%% file: paper.tex
\documentclass[journal]{IEEEtran}
\input{preamble.tex}
\def\arxiv{}

\newcommand\copyrighttext{
    \scriptsize
    \copyright{} 2022 IEEE. Personal use of this material is permitted.
    Permission from IEEE must be obtained for all other uses, in any current or
    \\ future media, including reprinting/republishing this material for
    advertising or promotional purposes, creating new collective works,\\ for
    resale or redistribution to servers or lists, or reuse of any copyrighted
    component of this work in other works.
}

\title{Magnetic Field Energy Harvesting in Railway}

\author{
    \IEEEauthorblockN{%
    Asbjørn~Engmark~Espe \orcidicon{0000-0003-1687-889X},
    \IEEEmembership{Member,~IEEE},\\
    Thomas~S.~Haugan, and Geir~Mathisen}

    \thanks{A. E. Espe and G. Mathisen are with the Department of Engineering
    Cybernetics, Norwegian University of Science and Technology, Trondheim,
    Norway (e-mail: asbjorn.e.espe@ntnu.no; geir.mathisen@ntnu.no)}
    \thanks{T. S. Haugan is with the Department of Electric Power Engineering,
    Norwegian University of Science and Technology, Trondheim, Norway (e-mail:
    thomas.haugan@ntnu.no)}
}
\IEEEtitleabstractindextext{%
    \input{0_abstract.tex}

\ifdefined\arxiv%
\begin{tikzpicture}[remember picture,overlay]%
    \node[rectangle, draw, minimum width=\textwidth, minimum height=24pt,%
    align=center, font=\scriptsize] at%
    ([yshift=-24pt]current page.north) {\copyrighttext};%
\end{tikzpicture}%
\fi
}

\makeatletter
\patchcmd{\@maketitle}
  {\addvspace{0.5\baselineskip}\egroup}
  {\addvspace{-1\baselineskip}\egroup}
  {}
  {}
\makeatother

\begin{document}
\maketitle

\thispagestyle{empty}

\IEEEdisplaynontitleabstractindextext

\input{1_introduction.tex}
\input{2_background.tex}
\input{3_modelling.tex}
\input{4_design.tex}
\input{5_lab_results.tex}

\input{6_rw_results.tex}

\input{7_conclusion.tex}

\bibliographystyle{IEEEtran}
\bibliography{IEEEabrv,paper}


\end{document}

%% file: preamble.tex
\usepackage[utf8]{inputenc}
\usepackage{graphicx}

\usepackage{tikz}
\usepackage{pgfplots}
\pgfplotsset{compat=1.3}
\usepackage{subfig}

\usepackage{amsmath}
\usepackage{xfrac}  
\DeclareUnicodeCharacter{2212}{-}

\usepackage[noadjust]{cite}
\usepackage[bookmarks=false]{hyperref}
\usepackage{cleveref}

\usepackage{booktabs}
\usepackage[
    binary-units,
    detect-all,
    text-ohm = $\Omega$
]{siunitx}

\usepackage{scalerel}

\Crefformat{equation}{(#2#1#3)}
\Crefrangeformat{equation}{(#3#1#4) to~(#5#2#6)}
\Crefmultiformat{equation}{(#2#1#3)}{ and~(#2#1#3)}{, (#2#1#3)}{, and~(#2#1#3)}
\Crefname{equation}{Equation}{Equations}
\Crefname{figure}{Figure}{Figures}
\Crefname{tabular}{Table}{Tables}

\newcommand{\etal}{\textit{et al.}\@}

\newcommand{\norrailfreq}{\num{16}\,\sfrac{2}{3}\,\si{\hertz}}

\definecolor{gray2}{gray}{0.2}
\definecolor{gray5}{gray}{0.5}
\definecolor{gray7}{gray}{0.7}
\definecolor{gray9}{gray}{0.9}
\definecolor{matlab1}{rgb}{0, 0.4470, 0.7410}
\definecolor{matlab2}{rgb}{0.8500, 0.3250, 0.0980}
\definecolor{matlab3}{rgb}{0.9290, 0.6940, 0.1250}
\definecolor{matlab4}{rgb}{0.4940, 0.1840, 0.5560}
\definecolor{matlab5}{rgb}{0.4660, 0.6740, 0.1880}
\definecolor{matlab6}{rgb}{0.3010, 0.7450, 0.9330}
\definecolor{matlab7}{rgb}{0.6350, 0.0780, 0.1840}
\colorlet{pgfblue}{matlab1}
\colorlet{pgfred}{matlab2}
\colorlet{magcol}{gray5}

\DeclareSIUnit\permille{\text{\textperthousand}}
\DeclareSIUnit\mAh{mAh}
\DeclareSIUnit\kph{km/h}

\usetikzlibrary{
    calc,
    patterns,
    arrows.meta,
    intersections,
    positioning,
    decorations.pathreplacing,
    decorations.markings
}

\tikzset{>=Stealth}

\newcommand\centerarrow[1][]{{\arrow[xshift={1.5pt + 2.25\pgflinewidth}, #1]{>}}}
\tikzstyle{circmagfield} = [
    color=magcol,
    decoration={markings,
        mark=at position 0 with {\centerarrow},
        mark=at position 0.33 with {\centerarrow},
        mark=at position 0.66 with {\centerarrow},
    },
    postaction={decorate}
]

\tikzstyle{strmagfield} = [
    color=magcol,
    decoration={markings,
        mark=at position 15pt with {\centerarrow},
        mark=at position 1 - 15pt with {\centerarrow},
    },
    postaction={decorate}
]

\pgfplotsset{
    legend image with text/.style={
        legend image code/.code={%
            \node[anchor=center] at (0.3cm, 0cm) {#1};
        }
    },
    model/.style={
        dashed, thick, domain=20:210
    },
    measured/.style={
        only marks, mark options={scale=0.75}
    },
    measured large/.style={
        only marks, mark options={scale=1.0}
    },
    coil a/.style={
        color=pgfblue, forget plot
    },
    coil b/.style={
        color=pgfred, forget plot
    },
}%

\usetikzlibrary{svg.path}

\definecolor{orcidlogocol}{HTML}{A6CE39}
\tikzset{
  orcidlogo/.pic={
    \fill[orcidlogocol] svg{M256,128c0,70.7-57.3,128-128,128C57.3,256,0,198.7,0,128C0,57.3,57.3,0,128,0C198.7,0,256,57.3,256,128z};
    \fill[white] svg{M86.3,186.2H70.9V79.1h15.4v48.4V186.2z}
                 svg{M108.9,79.1h41.6c39.6,0,57,28.3,57,53.6c0,27.5-21.5,53.6-56.8,53.6h-41.8V79.1z M124.3,172.4h24.5c34.9,0,42.9-26.5,42.9-39.7c0-21.5-13.7-39.7-43.7-39.7h-23.7V172.4z}
                 svg{M88.7,56.8c0,5.5-4.5,10.1-10.1,10.1c-5.6,0-10.1-4.6-10.1-10.1c0-5.6,4.5-10.1,10.1-10.1C84.2,46.7,88.7,51.3,88.7,56.8z};
  }
}

\newcommand\orcidicon[1]{\href{https://orcid.org/#1}{\raisebox{1pt}{\scalerel*{
\begin{tikzpicture}[yscale=-1,transform shape]
\pic{orcidlogo};
\end{tikzpicture}
}{|}}}}

%% file: 0_abstract.tex
\begin{abstract}
Magnetic field energy harvesting (MFEH) is a method by which a system can
harness an ambient, alternating magnetic field in order to scavenge energy.
Presented in this article is a novel application of the concept aimed at the
magnetic fields surrounding the rail current in electrified railway. Due to its
non-invasive nature, the approach has the potential to be widely deployed as
part of low-cost trackside condition monitoring systems in order to increase
lifetime and reduce maintenance requirements. In this work, the viability of
MFEH in railway is substantiated experimentally---two different configurations
are assessed both in a controlled laboratory environment, as well as in situ
along Norwegian railway. When placed near an emulated section of railway
carrying \SI{200}{\ampere} in the laboratory, the power output of the system is
up to \SI{40.5}{\milli\watt} at \SI{50}{\hertz} and \SI{4.15}{\milli\watt} at
\norrailfreq{}. In the field, the prototype system harvests
\SI{109}{\milli\joule} from a single freight train passing by, rendering an
estimated daily energy output of \SI{1.14}{\joule} in a moderately-trafficked
location. It is argued that the approach could indeed eliminate the need for
battery replacements, and potentially increase the lifetime of an
energy-efficient, battery-powered condition monitoring system indefinitely.
\end{abstract}

%% file: 1_introduction.tex
\section{Introduction}
\IEEEPARstart{A}{} substantial number of both commercial and private actors
routinely rely on the rail networks for the timely and secure transportation of
goods and passengers as part of their daily endeavours. In 2016, over 440
billion tonne-kilometres and 470 billion passenger-kilometres were recorded
across Europe \cite{eu18}. Accordingly, the safe and continuous operation of
the related infrastructure are of cardinal significance in our modern society,
and hence fundamental priorities for railway administrations. In Norway alone,
there are more than \num{2600} bridges and \num{700} tunnels that must be kept
available and properly maintained at all times \cite{jbd18}. Indeed, as the
frequency and severity of extreme weather appear to increase in response to
climate change, a considerable amount of resources is expended to support
traditional periodic maintenance schemes.

Enabled by advances in smart maintenance technologies, condition-based and
predictive maintenance schemes have been introduced in a diverse range of areas
as a more resource-conservative approach \cite{akkermans16}. In the last few
years, railway authorities have embodied this paradigm shift by increasingly
taking interest in smart maintenance and its enabling technologies
\cite{banenor17,takikawa16}. As a central element of smart maintenance in
railway, trackside condition monitoring systems may be installed to monitor the
structural integrity of infrastructure such as bridges, tracks, turnouts, and
tunnels \cite{hodge15}. Due to their low cost, low energy consumption,
flexibility, and rapid deployment, wireless sensor networks (WSNs) have become
an attractive solution for this type of real-time condition monitoring
\cite{akyildiz02}.

Traditionally powered by batteries, wireless sensor nodes are naturally limited
by the capacity of their energy reserves. While such a node typically requires
tens to hundreds of microwatts in active operation, and several hundred
milliwatts during wireless transmission, the node may substantially decrease
average power usage by spending most of its time in ultra-low power modes
\cite{boyle16}. To ensure continuous operation, an additional maintenance step
is in many cases stipulated in the form of periodic battery replacements, which
is generally undesirable for end users \cite{gjerstad20}. As an alternative,
energy harvesting can be employed to regularly charge the system and thereby
increase its battery-life indefinitely, provided that the amount of harvested
energy is sufficient \cite{adumanu18}.

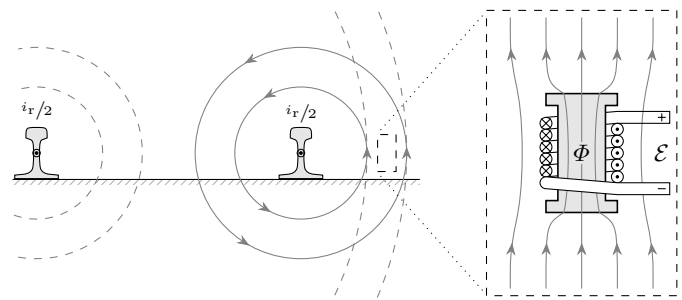
\begin{figure}[!b]
    \centering
    \vspace{-10pt}
    \input{railway_mfeh.tex}
    \caption{An alternating current $i_r$ carried by the rails gives
    rise to a magnetic field from which energy can be harvested. A
    solid-core wire coil placed nearby will experience a varying flux $\varPhi$
    through its loops, inducing an electromotive force $\mathcal{E}$.} 
    \label{fig:railway_mfeh}
\end{figure}

Many conventional energy harvesting sources such as solar energy, wind energy,
and vibration energy have been successfully employed for railway condition
monitoring systems \cite{ulianov20}. Interestingly, while magnetic field energy
harvesting (MFEH) has emerged as an attractive approach for energy harvesting
near overhead power lines \cite{yang20}, the technique has received minimal
recognition in a railway context, in spite of the fact that the contact line
system of electrified railway closely resembles power lines in many aspects. In
fact, the only previous work investigating the technique in this context
appears to be the recent research by Kuang \etal{} \cite{kuang21}, with results
purely from a laboratory setting. \Cref{fig:railway_mfeh} illustrates a
particularly promising variant of the concept that is investigated in this
work. As trains pass by, current is drawn from the contact line and returned
through the rails, giving rise to a magnetic field. In AC systems, the
time-varying nature of this field allows it to be a source for energy
harvesting if an electromagnetic coil is placed in its vicinity.

The novelty of this work lies in its pioneering in situ demonstration of MFEH
for trackside condition monitoring systems in railway, substantiated by
laboratory experiments and a fairly accurate theoretical model. While no
real-world demonstration of this energy harvesting approach can be found in the
literature, its potential is supported by the modelling and laboratory work of
Kuang \etal{} \cite{kuang21}, as well as a preliminary feasibility study by the
authors of this article which indicates that MFEH may indeed be a feasible
approach for railway \cite{espe20}. The latter study is of a purely theoretical
nature and thus does not provide any laboratory or real-world results. In
\cite{kuang21}, an energy harvester design to be placed underneath the railway
tracks is introduced. Being extremely close to---and partially enclosing---the
rail, their design is in a laboratory environment able to harvest
\SI{200}{\milli\watt} and \SI{5}{\watt} with a rail current magnitude of
\SI{100}{\ampere} and \SI{520}{\ampere}, respectively. However, the study
provides no in situ results and only considers a single rail. In addition, it
lacks simulated or experimental results for the electrification frequency of
\norrailfreq{}, making its results of limited value in countries employing this
frequency.

Particularly relevant solutions from a power grid context are free-standing
ferrite-core energy harvesting coils to be placed in the proximity of overhead
power lines, such as the works by Yuan \etal{} \cite{yuan15, yuan17}. The
former presents a ferrite core in the shape of a bow-tie, reporting a power
output of \SI{360}{\micro\watt} from a \SI{7}{\micro\tesla} magnetic field,
while the latter improves this figure fourfold introducing a more complex
helical design. In \cite{jiang16}, a small device physically attached to the
power line is able to harvest more than \SI{30}{\milli\watt}. At the expense of
a more involved installation procedure, \cite{wu18} reports a figure of
\SI{230}{\milli\watt} using a flux guide that wraps around the power line
itself.

While MFEH is a novel approach within railway, many other energy harvesting
variants have been demonstrated previously \cite{ulianov20}. Inductive coils
have been physically attached to the rails in order to harness the mechanical
energy stemming from vertical displacement, for instance in the work by
Pourghodrat \etal{} \cite{pourghodrat14}. Wischke \etal{} \cite{wischke11}
examined the use of piezoelectric vibration energy harvesters. In
\cite{greenrail17}, photovoltaic panels were attached to railway sleepers,
while Nandan \etal{} \cite{nandan17} and Pan \etal{} \cite{pan19} both
introduce custom wind turbine designs.

%% file: railway_mfeh.tex
\begin{tikzpicture}
    \path[draw, fill=gray9,
        x=0.11pt, y=0.11pt] (0, -88) -- (-75, -88) --
        (-75, -76) -- (-45, -73) to[out=5, in=260] (-13, -50) .. controls
        (-10, -40) and (-8, 20) .. (-13, 37) to[out=100, in=340]
        (-23, 47) -- (-36.5, 52) .. controls(-36.5, 88) .. (0, 88) ..
        controls (36.5, 88) .. (36.5, 52) -- (23, 47) to[out=190, in=80]
        (13, 37) .. controls (8, 20) and (10, -40) .. (13, -50)
        to[out=280, in=175] (45, -73) -- (75, -76) -- (75, -88) -- cycle;

    \path[draw, fill=gray9,
        x=0.11pt, y=0.11pt, xshift=-100pt] (0, -88) -- (-75, -88) --
        (-75, -76) -- (-45, -73) to[out=5, in=260] (-13, -50) .. controls
        (-10, -40) and (-8, 20) .. (-13, 37) to[out=100, in=340]
        (-23, 47) -- (-36.5, 52) .. controls(-36.5, 88) .. (0, 88) ..
        controls (36.5, 88) .. (36.5, 52) -- (23, 47) to[out=190, in=80]
        (13, 37) .. controls (8, 20) and (10, -40) .. (13, -50)
        to[out=280, in=175] (45, -73) -- (75, -76) -- (75, -88) -- cycle;

    \path[pattern=north east lines, pattern color=gray7] (-109pt, -10pt) --
    (45pt, -10pt) -- ++(0, -2pt) -- (-109pt, -12pt) -- cycle;
    \draw (-109pt, -10pt) -- (45pt, -10pt);

    \node[circle, inner sep=0.1pt, draw, fill] at (0, 0) {};
    \node[circle, inner sep=1pt, draw] at (0, 0) {};
    \node[circle, inner sep=0.1pt, draw, fill] at (-100pt, 0) {};
    \node[circle, inner sep=1pt, draw] at (-100pt, 0) {};

    \draw[circmagfield] (0, 0) circle (25pt);
    \draw[circmagfield] (0, 0) circle (40pt);
    \draw[color=magcol, dashed] (-100pt, 0) ++(-110:25pt) arc (-110:110:25pt);
    \draw[color=magcol, dashed] (-100pt, 0) ++(-100:40pt) arc (-100:100:40pt);
    \draw[color=magcol, dashed] (-100pt, 0) ++(-25.5:125pt) arc (-25.5:25.5:125pt);
    \draw[color=magcol, dashed] (-100pt, 0) ++(-23:140pt) arc (-23:23:140pt);

    \node[draw, dashed, minimum height=14pt, minimum width=7pt] (focus) at
        (32.5pt, 0) {};
    \coordinate[right=70pt of focus] (zoom);

    \begin{scope}[shift={(zoom.center)}, scale=0.9]
        \draw[semithick, fill=gray9] (-15pt, -25pt) |- ++(5pt, 5pt) -- ++(0, 40pt) -|
        ++(-5pt, 5pt) -- ++(30pt, 0) |- ++(-5pt, -5pt) -- ++(0, -40pt) -| ++(5pt,
        -5pt) -- cycle;
        \draw[dashed] (zoom) ++(-40pt, -60pt) coordinate (zoom_bot) --
        ++(0, 120pt) coordinate (zoom_top) -- ++(80pt, 0) |- cycle;
        \draw[strmagfield] (-30pt, -57pt) -- (-30pt, -50pt)
            .. controls (-30pt, -30pt) and (-25pt, -30pt) .. (-25pt, 0)
            .. controls (-25pt, 30pt) and (-30pt, 30pt) .. (-30pt, 50pt) --
            (-30pt, 57pt);
        \draw[strmagfield] (-15pt, -57pt)
            .. controls (-15pt, -30pt) .. (-7.5pt, -25pt)
            .. controls (-5pt, -15pt) and (-5pt, 15pt) .. (-7.5pt, 25pt)
            .. controls (-15pt, 30pt) .. (-15pt, 57pt);
        \draw[strmagfield] (0, -57pt) -- (0, 57pt);
        \draw[strmagfield] (15pt, -57pt)
            .. controls (15pt, -30pt) .. (7.5pt, -25pt)
            .. controls (5pt, -15pt) and (5pt, 15pt) .. (7.5pt, 25pt)
            .. controls (15pt, 30pt) .. (15pt, 57pt);
        \draw[strmagfield] (30pt, -57pt) -- (30pt, -50pt)
            .. controls (30pt, -30pt) and (25pt, -30pt) .. (25pt, 0)
            .. controls (25pt, 30pt) and (30pt, 30pt) .. (30pt, 50pt) --
            (30pt, 57pt);
    \end{scope}

    \begin{scope}[shift={(zoom.center)}, scale=0.9]
        \foreach \y in {0,...,4}{
            \draw (-15pt, \y*5pt - 7.5pt) circle (2.5pt);
            \draw (-15pt, \y*5pt - 5pt) -- ++(5pt, 0.416pt);
            \draw (15pt, \y*5pt - 10pt) circle (2.5pt);
            \draw (15pt, \y*5pt - 7.5pt) -- ++(-5pt, -0.416pt);

            \draw[thin] (-15pt, \y*5pt - 7.5pt) ++(-2pt, -2pt) -- ++(4pt, 4pt)
            ++(-4pt, 0) -- ++(4pt, -4pt);
            \draw[fill=black] (15pt, \y*5pt - 10pt) circle (0.5pt);
        }

        \draw (-15pt, -10pt) -- ++(5pt, 0.416pt);
        \draw[fill=white] (37pt, -12.5pt) -- coordinate[pos=0.5, xshift=-3pt] (neg)
        ++(0, -5pt) -- ++(-22pt, 0) -- ++(-30pt, 2.5pt) arc(-90:-270:2.5pt) --
        ++(30pt, -2.5pt) -- cycle;
        \draw (10pt, 17.5pt) ++(0, -0.416pt) -- ++(5pt, 0.416pt) -- ++(22pt, 0)
        -- coordinate[pos=0.5, xshift=-3pt] (pos) ++(0, -5pt) -- ++(-22pt, 0);
    \end{scope}

    \draw[dotted] ($(focus.north west)!2pt!(zoom_top)$)--
        ($(zoom_top)!2pt!(focus.north east)$);
    \draw[dotted] ($(focus.south west)!2pt!(zoom_bot)$)--
        ($(zoom_bot)!2pt!(focus.south east)$);

    \draw (pos) ++(-1.5pt, 0) -- ++(3pt, 0) ++(-1.5pt, -1.5pt) -- ++(0, 3pt);
    \draw (neg) ++(-1.5pt, 0) -- ++(3pt, 0);
    \path (pos) -- node[pos=0.5] {\small $\mathcal{E}$} (neg);

    \node[above, yshift=9pt] at (-100pt, 0) {\scriptsize $\sfrac{i_\mathrm{r}}{2}$}; 
    \node[above, yshift=8pt] at (0, 0) {\scriptsize $\sfrac{i_\mathrm{r}}{2}$}; 
    \node[inner sep=1pt, fill=gray9] (phi) at (zoom)
        {\small $\varPhi$};
\end{tikzpicture}

%% file: 2_background.tex
\section{Background}\label{sec:background}
A railway electrification system may employ either direct current (DC) or
alternating current (AC) for power distribution. As of 2018, \SI{63}{\percent}
of the world's electrified railway uses AC \cite{kiessling18}. The most common
voltage system is \SI{25}{\kilo\volt}, \SI{50}{\hertz}, while a handful of
nations in Central Europe and Scandinavia use \SI{15}{\kilo\volt},
\norrailfreq{}\footnote{In 2000, the nominal frequency was changed to
\SI{16.7}{\hertz} in Germany, Austria, and Switzerland.}. For practical
reasons, the in situ system presented in this work targets Norwegian railway
which employs the latter. Nonetheless, it is established by laboratory
experiments that a frequency of \SI{50}{\hertz} leads to a significant
improvement in output power over \norrailfreq{}, making the solution
particularly attractive at this higher frequency.

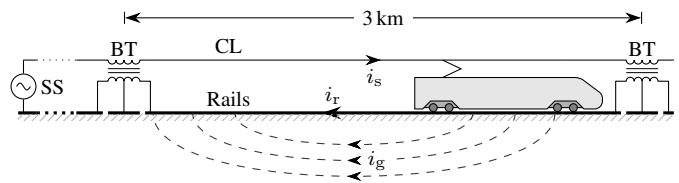
\begin{figure}
    \centering
    \input{system_b.tex}
    \vspace{-15pt}
    \caption{\textit{System B}---the most common electrification configuration
    in Norway. Power is delivered from the nearest substation (SS) to the
    locomotive through the contact line circuit. The current $i_\mathrm{s}$
    carried by the contact line (CL) is complemented by currents through the
    rails ($i_\mathrm{r}$) and ground ($i_\mathrm{g}$). At regular intervals,
    booster transformers (BT) are employed to minimise the ground leakage
    currents.}
    \label{fig:system_b}
\end{figure}

A substantial part of electrified railway in Norway employs the configuration
known as \textit{System B} \cite{jbv97}, illustrated in \Cref{fig:system_b}.
Railway conforming to this standard generally lacks a dedicated return
conductor and rather utilises the rails themselves to complete the electric
circuit; current drawn from locomotives is mainly carried by the rails back to
the nearest substation. New and renovated railway in Norway is typically built
to more modern standards---\textit{System C} or \textit{D}---which do feature
dedicated return conductors. However, even in these configurations, current
will still flow through the rails a short distance from the locomotive to the
nearest pole.

The hypothesis in this work is that the current in the rails is of such a
magnitude that it could serve as an energy harvesting source by magnetic
induction. As mentioned in the previous, similar studies in a power grid
context demonstrate the viability of using MFEH to supplement battery-powered
systems. If compared to overhead power lines, railway tracks are generally of a
more accessible nature, allowing energy harvesting solutions to be placed very
close to the current without necessitating expensive procedures or invasive
structures. Since the magnitude of the magnetic fields established by the
currents scale inversely with distance, this renders MFEH a promising approach
for energy harvesting in railway. A drawback compared to power lines is the
intermittency of the current. For modern railway, a substantial current is
carried by the rails only when an electric locomotive passes by. In the case of
\textit{System B} and similar configurations, the approach is more attractive
as current passes through the rails for a significantly longer period of time.

In contrast to alternative energy harvesting approaches that depend on
environmental conditions beyond control---such as photovoltaic panels or wind
turbines---a MFEH system relies upon a considerably more predictable source of
energy. Railway traffic is often strictly scheduled, making it possible to
quite accurately predict future energy output from a MFEH system. If used in
conjunction with battery-powered trackside condition monitoring systems, this
desirable trait could improve the confidence level of power management
optimisation techniques \cite{ahmed19}. Compared to mechanical energy
harvesters in railway, an MFEH approach is both galvanically and physically
isolated from the rails, and generally offers a cheaper and less invasive
installation procedure.

%% file: system_b.tex
\begin{tikzpicture}[>={Stealth[width=4pt, length=6pt]}]
    \node[circle, draw, minimum size=8pt] (source) at (-120pt, 10pt) {};
    \begin{scope}[shift={(source)}]
        \draw[x=1.3pt,y=1.3pt] (-2, 0) sin (-1, 1) cos (0, 0) sin (1, -1) cos (2, 0);
    \end{scope}

    \coordinate (rail_begin) at (source |- 0, 0);
    \coordinate (cl_begin) at (source |- 0, 20pt);
    \draw (source) -- (rail_begin);

    \draw[line width=1.2pt,
        decoration={markings, mark=at position 0.55 with
        {\centerarrow[scale=1.2]}}]
        (rail_begin) ++(-2pt, 0) coordinate (ground_begin) -- ++(7pt, 0)
        edge[dotted] ++(15pt, 0) ++(15pt, 0) -- ++(10pt, 0) ++(2pt, 0) --
        ++(12pt, 0) ++(2pt, 0) coordinate (r_start)
        edge[draw=none] coordinate[pos=0.2] (raillabel) ++(158pt, 0)
        ++(158pt, 0) coordinate (r_end)
        edge[postaction={decorate}] coordinate[pos=0.55] (i_r) ++(-158pt, 0)
        -- ++(22pt, 0) ++(2pt, 0) -- ++(12pt, 0) ++(2pt, 0) -- ++(4pt, 0) coordinate (ground_end);

    \path[pattern=north east lines, pattern color=gray7] ($(ground_begin)+(0, -0.75pt)$) --
    ($(ground_end)+(0, -0.75pt)$) -- ++(0, -2pt) -| cycle;

    \draw[line join=bevel,
        decoration={markings, mark=at position 0.55 with {\centerarrow}}]
        (source) -- (cl_begin) -- ++(5pt, 0)
        edge[dotted] ++(15pt, 0) ++(15pt, 0) -- ++(12pt, 0)
        arc(-180:0:1.5pt) arc(-180:0:1.5pt)
        arc(-180:0:1.5pt) arc(-180:0:1.5pt) -- ++(2pt, 0)
        edge[postaction={decorate}] coordinate[pos=0.55] (i_s) ++(158pt, 0)
        edge[draw=none] coordinate[pos=0.2] (cllabel) ++(158pt, 0)
        ++(158pt, 0) -- ++(24pt, 0)
        arc(-180:0:1.5pt) arc(-180:0:1.5pt)
        arc(-180:0:1.5pt) arc(-180:0:1.5pt)
        -- ++(2pt, 0) -- ++(4pt, 0);

    \draw[line join=bevel] (rail_begin) ++(28pt, 0) |- ++(4pt, 12pt)
        ++(0, 3.25pt) edge[very thin] ++(12pt, 0)
        ++(0, 1.5pt) edge[very thin] ++(12pt, 0) ++(0, -4.75pt)
        arc(180:0:1.5pt) arc(180:0:1.5pt) edge ++(0, -12pt) coordinate
        (booster1) arc(180:0:1.5pt) arc(180:0:1.5pt)
        -| ++(4pt, -12pt)

        ++(176pt, 0) |- ++(4pt, 12pt)
        ++(0, 3.25pt) edge[very thin] ++(12pt, 0)
        ++(0, 1.5pt) edge[very thin] ++(12pt, 0) ++(0, -4.75pt)
        arc(180:0:1.5pt) arc(180:0:1.5pt) edge ++(0, -12pt) coordinate
        (booster2) arc(180:0:1.5pt) arc(180:0:1.5pt)
        -| ++(4pt, -12pt);

    \node[above] at (i_r) {\footnotesize $i_\mathrm{r}$}; 
    \node[below] at (i_s) {\footnotesize $i_\mathrm{s}$}; 
    \node[above, yshift=-1pt] at (cllabel) {\footnotesize CL};
    \node[above, yshift=-1pt] at (raillabel) {\footnotesize Rails};
    \node[right, xshift=3pt] at (source) {\footnotesize SS};

    \node[above, yshift=6pt] at (booster1) {\footnotesize BT};
    \node[above, yshift=6pt] at (booster2) {\footnotesize BT};

    \draw[|<->|] (booster1 |- 0, 36pt) -- (booster2 |- 0, 36pt)
        node[midway, fill=white] {\footnotesize \SI{3}{\kilo\meter}};

    \begin{scope}[every path/.style={
        decoration={markings, mark=at position 0.5 with {\centerarrow[scale=0.85]}}
        }
    ]
        \draw[dashed, gray2] ($(r_end) - (3pt, 0)$)
            .. controls ++(0, -32pt) and ++(0, -32pt) ..
            ($(r_start) + (3pt, 0)$);
        \draw[dashed, gray2] ($(r_end) - (18pt, 0)$)
            .. controls ++(0, -24pt) and ++(0, -24pt) ..
            coordinate[midway] (i_leak) ($(r_start) + (18pt, 0)$);
        \draw[dashed, gray2] ($(r_end) - (34pt, 0)$)
            .. controls ++(0, -16pt) and ++(0, -16pt) ..
            ($(r_start) + (34pt, 0)$);
        \path[postaction={decorate}] ($(r_end) - (3pt, 0)$)
            .. controls ++(0, -32pt) and ++(0, -32pt) ..
            ($(r_start) + (3pt, 0)$);
        \path[postaction={decorate}] ($(r_end) - (18pt, 0)$)
            .. controls ++(0, -24pt) and ++(0, -24pt) ..
            ($(r_start) + (18pt, 0)$);
        \path[postaction={decorate}] ($(r_end) - (34pt, 0)$)
            .. controls ++(0, -16pt) and ++(0, -16pt) ..
            ($(r_start) + (34pt, 0)$);
        \node[right, black, fill=white, inner sep=1pt, xshift=4pt] at (i_leak)
            {\footnotesize $i_\mathrm{g}$}; 
    \end{scope}

    \coordinate (loc) at ($(r_end) + (4pt, 2pt)$);
    \coordinate (frwh) at ($(loc) - (3pt, 0)$);
    \coordinate (bawh) at ($(loc) - (50pt, 0)$);
    \draw[fill=gray7] (frwh) -| ++(8pt, 4pt) -| ++(-16pt, -4pt) -- cycle;
    \draw[fill=gray7] (bawh) -| ++(8pt, 4pt) -| ++(-16pt, -4pt) -- cycle;
    \draw[fill=gray9] (frwh) ++(0, 2.5pt) -- ++(4pt, 0) -- ++(3pt, -3pt) --
        ++(4pt, 0) arc(-90:0:3pt) -- ++(0, 1pt) arc(0:60:3pt) --
        ++(-6pt, 4.5pt) arc(60:90:6pt)
        edge[draw=none] coordinate[pos=1] (panto) ++(-50pt, 0)
        -| ($(bawh) + (-10pt, -0.5pt)$) -- ($(bawh) + (-7pt, -0.5pt)$) --
        ++(3pt, 3pt) -- ++(8pt, 0) -- ++(3pt, -3pt) --
        ($(frwh) + (-7pt, -0.5pt)$) -- ++(3pt, 3pt) -- cycle;
    \draw[fill=gray5] (frwh) ++(3pt, 0) circle (1.5pt);
    \draw[fill=gray5] (frwh) ++(-3pt, 0) circle (1.5pt);
    \draw[fill=gray5] (bawh) ++(3pt, 0) circle (1.5pt);
    \draw[fill=gray5] (bawh) ++(-3pt, 0) circle (1.5pt);
    \coordinate (panto_end) at (cl_begin -| panto);
    \coordinate (panto_mid) at ($(panto) !.5! (panto_end) + (7pt, 0)$);
    \draw (panto) -- (panto_mid) -- (panto_end);
\end{tikzpicture}

%% file: 3_modelling.tex
\section{Mathematical modelling}
\begin{figure}
    \centering
    \subfloat[\label{fig:sim_scenario}Simulation scenario]{%
        \includegraphics[width=\columnwidth]{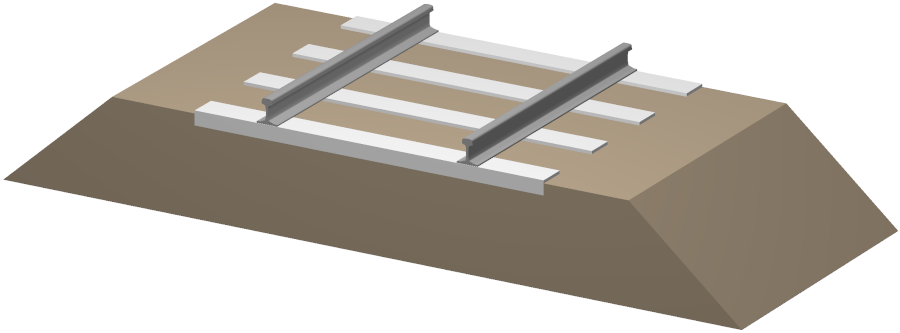}
    }

    \subfloat[\label{fig:magn_flux_sim}Simulated magnetic flux density
    (\textit{CST Studio 2019})]{%
        \input{b_field.pgf}
    }
    \caption{A typical scenario is illustrated in
    \protect\subref{fig:sim_scenario}, consisting of \textit{60E1}-type rails,
    concrete sleepers, and granite ballast. A cross section of the magnetic flux
    density field as a current of \SI{50}{\ampere} RMS at \norrailfreq{} is
    carried by each rail is shown in \protect\subref{fig:magn_flux_sim}.}
\end{figure}

Compared to an overhead power line, the railway scenario is particularly
challenging to model mathematically due to the complex shapes involved. However,
certain assumptions and simplifications can make the problem less complicated at
the expense of some accuracy. As an initial effort to aid in modelling, a
typical scenario---depicted in \Cref{fig:sim_scenario}---was simulated, with the
resulting magnetic flux density shown in \Cref{fig:magn_flux_sim}.

\subsection{Modelling assumptions}
There are three main assumptions in the following model. A preliminary
simplification is to treat the rails as two infinitely long, straight, and
parallel conductors, and thereby assume any curvature to be negligible compared
to the dimensions of the energy harvesting device. This simplification is
generally valid for a real-world scenario, as the minimum curve radius of
low-speed railway is typically at least \SI{400}{\meter} \cite{lee20}---several
orders of magnitude larger than the distance to the harvester. Furthermore, as
demonstrated in \cite{espe20}, beyond a radial distance of about
\SI{0.4}{\meter} the rail shape can be approximated as a cylinder with
negligible loss of accuracy. Lastly, the environment of the rails is to be
modelled as vacuum, as the permeabilities of steel and the MFEH core are far
greater than that of surrounding materials such as air, wood, concrete, and
crushed rock ballast, all of which have relative permeabilities close to 1. The
recommended material for track ballast in Norway is crushed granite, the
susceptibility of which is reported to be between \num{1.26e-5} and
\num{5.03e-2} \cite{carmichael82}---meaning a magnetic permeability comparable
to that of air.

\subsection{Magnetic field}\label{sec:mag_field}
With these assumptions, the Biot-Savart law \cite{jiles16} is used to describe
the ambient magnetic field strength around the rails,
\begin{equation}\label{eq:biot_savart}
    \mathbf{H} = \frac{\sfrac{i_\mathrm{r}}{2}}{2 \pi r_\mathrm{f}}
    \hat{\boldsymbol{\phi}}_\mathrm{f} + \frac{\sfrac{i_\mathrm{r}}{2}}{2 \pi
    r_\mathrm{n}} \hat{\boldsymbol{\phi}}_\mathrm{n} \ ,
\end{equation}
where $i_\mathrm{r}$ is the rail current, assumed to be distributed evenly
between the rails. Cylindrical co-ordinates are employed, meaning
$r_\mathrm{f}$, $\hat{\boldsymbol{\phi}}_\mathrm{f}$, $r_\mathrm{n}$, and
$\hat{\boldsymbol{\phi}}_\mathrm{n}$ are the radii and unit vectors in the
rotational direction for the far and near rail, respectively.

Assuming a physical configuration similar to the one illustrated in
\Cref{fig:railway_mfeh}, the far and near rail are named according to their
distance from the potential placement of the energy harvester. Furthermore, it
is assumed that both rails are vertically centred on the same horisontal plane
at $y = 0$, and that they are separated by a constant distance of
$d_\mathrm{rr}$. With the exception of the area between the rails, the
contribution from each rail to the total magnetic field strength is parallel
and additive on this plane, making it a promising region for placement of the
energy harvester. Specifically, at any point in this region, the unit vectors
$\hat{\boldsymbol{\phi}}_\mathrm{f}$ and $\hat{\boldsymbol{\phi}}_\mathrm{n}$
are equal to $\hat{\boldsymbol{y}}$, and the relation $r_\mathrm{f} =
r_\mathrm{n} + d_\mathrm{rr}$ holds. The magnetic field strength can therefore
be expressed
\begin{equation}\label{eq:biot_savart2}
    \mathbf{H}_0 = \left( \frac{i_\mathrm{r}}{4 \pi \left(r_\mathrm{n} +
    d_\mathrm{rr}\right)} + \frac{i_\mathrm{r}}{4 \pi r_\mathrm{n}} \right)
    \hat{\boldsymbol{y}} = \frac{i_\mathrm{r}}{2 \pi r_\mathrm{e}} \hat{\boldsymbol{y}} \ ,
\end{equation}
in which the variable
\begin{equation}\label{eq:eff_dist}
    r_\mathrm{e} = \frac{2 r_\mathrm{n} \left(r_\mathrm{n} +
    d_\mathrm{rr}\right)}{2r_\mathrm{n} + d_\mathrm{rr}}
\end{equation}
is introduced as a measure of \textit{effective radius}.

The expression in \Cref{eq:biot_savart2} is used in the following to describe
the applied magnetic field strength for the real-world scenario, thereby
introducing an assumption that the magnetic field is close to uniform in the
vicinity of the energy harvesting device. This is undoubtedly an approximation
of its true shape, albeit necessary in order to reduce the complexity of the
mathematical model. Presumably, the modelling error introduced by this choice
will decrease as the distance $r_\mathrm{n}$ increases and the curvature of the
field is reduced.

Another aspect of the scenario that is not accounted for in the model above is
the alternating nature of the current, since the Biot-Savart law is only
applicable for a direct current. The displacement current, however, is
typically negligible for frequencies below the \si{\tera\hertz} range
\cite{jiles16}. In addition, the model does not consider the magnetic
properties of the rails themselves; as described by Lenz's law, eddy currents
induced in the rails will generally lead to a reduction of the magnetic field
strength. While it has been demonstrated that interaction of this kind between
the coil and the rail is generally negligible beyond distances as small as
\SI{0.19}{\meter} \cite{kuang21}, mutual inductance between the two rails is
not accounted for. In summary, the magnetic field simulated in
\Cref{fig:magn_flux_sim} deviates from the model presumably due to the
modelling approximations such as the conductor shape and nonlinearity in the
response to the AC frequency of \norrailfreq{}. Indeed, the simulated scenario
reports a flux density of \SI{22.2}{\micro\tesla} at a distance of
\SI{0.5}{\meter}, while the mathematical model yields a magnetic flux density
in free air of \SI{25.2}{\micro\tesla} at the same location.

\subsection{Loss mechanisms}
There are three main phenomena that will affect the efficiency of the energy
harvesting core itself: demagnetisation, hysteresis loss, and eddy current loss
\cite{jiles16}.

\subsubsection{Demagnetisation}
The energy harvesting device is constructed as a wire coiled around a solid core
with a relative permeability $\mu_\mathrm{r} > 1$. Since the core can not fully
enclose the rail, the generation of a magnetic dipole moment gives rise to a
demagnetising field opposing the applied field $\mathbf{H}_0$. The result is
that the magnetic field strength $\mathbf{H}_\mathrm{c}$ in the core is less
than $\mathbf{H}_0$, and may be expressed as \Cref{eq:demag_field}, where
$\mathbf{M}_\mathrm{c}$ is the core's magnetisation, and $N_\mathrm{d}$ is the
demagnetisation factor \cite{jiles16}.
\begin{equation}\label{eq:demag_field}
    \mathbf{H}_\mathrm{c} = \mathbf{H}_0 - N_\mathrm{d} \mathbf{M}_\mathrm{c}
\end{equation}

Under the assumption that the applied magnetic field is close to uniform and
weak to the extent that the core's response is linear, the magnetic flux density
in the core may be approximated by the scalar expression
\begin{equation}
    B_\mathrm{c} = \mu_\mathrm{e} \mu_0 H_0 \ ,
\end{equation}
where $\mu_\mathrm{e}$ is a measure of the core's \textit{effective}
permeability. A relation can be derived as shown in \Cref{eq:eff_mu}, revealing
the effective permeability to be dependent on the demagnetisation factor
$N_\mathrm{d}$, and residing in the range $1 \le \mu_\mathrm{e} \le
\mu_\mathrm{r}$ for $0 \leq N_\mathrm{d} \leq 1$.
\begin{equation}\label{eq:eff_mu}
    \mu_\mathrm{e} = \frac{\mu_\mathrm{r}}{1 + N_\mathrm{d}\left(\mu_\mathrm{r} -
    1\right)}
\end{equation}
$N_\mathrm{d}$ is mainly determined by core shape, and an effort should
therefore be made towards reducing it as much as possible, and thereby increase
the efficiency of the core.

\subsubsection{Hysteresis loss}
The hysteresis loss accounts for the work done by a periodically reversing
magnetic field in order to magnetise the core material. This type of loss
generally depends on the chosen material, as well as the frequency and
magnitudes of the applied magnetic field.

\subsubsection{Eddy current loss}
Eddy currents can have a great impact on the efficiency of an MFEH device
\cite{roscoe13}. The power loss is challenging to model accurately, however
\cite{jiles16} introduces the empirical model \Cref{eq:eddy_loss} for losses in
a cylinder stemming from eddy current, where $d$ is the diameter of the
cylinder, $f$ is the frequency, $B_\mathrm{p}$ is the peak magnetic flux
density, and $\rho$ is the resistivity of the core material.
\begin{equation}\label{eq:eddy_loss}
    W_\mathrm{ec} = \frac{\pi^2 B_\mathrm{p}^2 d^2 f^2}{16 \rho}
\end{equation}
From this equation, it is evident that a higher-resistivity material and
smaller-diameter core will help mitigate the losses. The losses also scale with both
magnetic flux density magnitude and frequency squared. It is therefore expected
that eddy currents will only have minor impact in this application due to the
low frequency and weak magnetic fields.

\subsection{Power output}
In the following, $\bar{I}$ is introduced as a shorthand for the
root-mean-square (RMS) phasor $I\angle\SI{0}{\degree}$, and in time-domain
equivalent to $i_\mathrm{r} = \sqrt{2} \, I \mathrm{cos}(\omega t)$. Likewise,
$\bar{B}_\mathrm{c}$, $\bar{H}_0$, and $\bar{V}_\mathrm{oc}$ denote RMS phasors
for the magnetic flux density, the magnetic field strength, and the
open-circuit voltage in the coil, respectively.

Faraday's law may be applied as shown in \Cref{eq:faraday} to determine the
induced electromotive force in the energy harvester. Inserting the applied
magnetic field strength from \Cref{eq:biot_savart2}, the relation between the
return current and open-circuit voltage is obtained.
\begin{equation}\label{eq:faraday}
    \bar{V}_\mathrm{oc} = N \frac{\mathrm{d}(\bar{B}_\mathrm{c}A)}{\mathrm{d}t}
    = N A \mu_\mathrm{e} \mu_0 j \omega \bar{H}_0
    = \frac{N A \mu_\mathrm{e} \mu_0}{2\pi r_\mathrm{e}} j \omega \bar{I}
\end{equation}
The angular frequency is denoted by $\omega$, and the number of turns and the
average enclosed area are given by $N$ and $A$, respectively.

The power $P_\ell$ dissipated in an impedance-matched load can then be
expressed as \Cref{eq:power_output}.
\begin{equation}\label{eq:power_output}
    P_\ell = \frac{\left|\bar{V}_\mathrm{oc}\right|^2}{4 R}
    = \frac{\left(N A \mu_\mathrm{e} \mu_0 \omega H_0 \right)^2}{4R}
    = \frac{\left(N A \mu_\mathrm{e} \mu_0 f I \right)^2}{4 R r_\mathrm{e}^2}
\end{equation}
In the above expression, $R$ denotes the coil's resistance, while $f =
\sfrac{\omega}{2 \pi}$ and $I$ are the frequency and RMS magnitude of the total
rail current that gives rise to the magnetic field.

%% file: 4_design.tex
\begin{figure}[t]
    \captionsetup[subfloat]{farskip=0pt}
    \subfloat[\label{fig:coila}]{%
        \includegraphics[width=0.4\columnwidth]{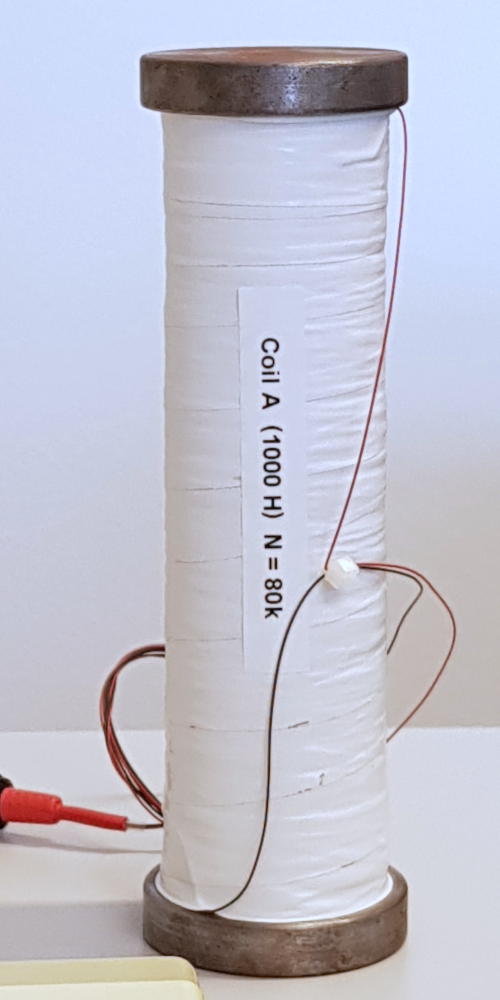}
    }
    \hfill
    \subfloat[\label{fig:coilb}]{%
        \includegraphics[width=0.4\columnwidth]{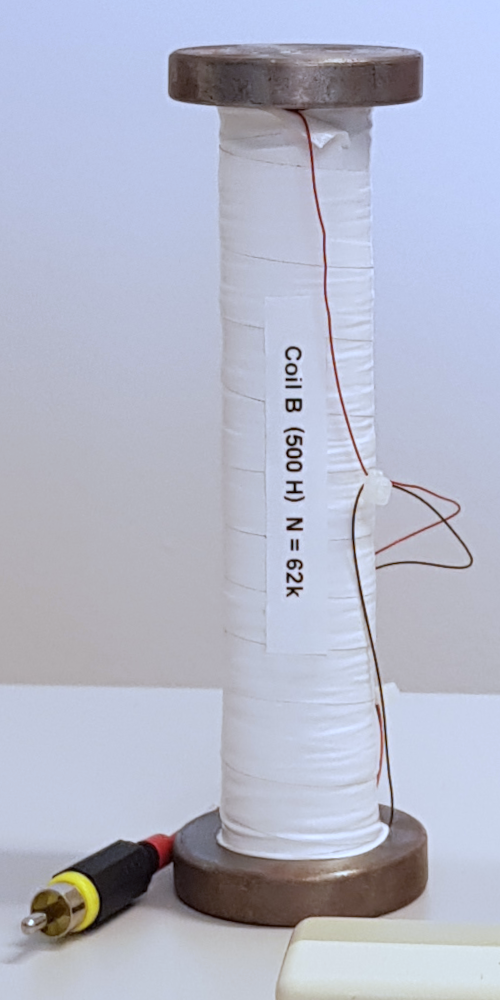}
    }
    \caption{Two coils have been constructed: \protect\subref{fig:coila}
    Coil A and \protect\subref{fig:coilb} Coil B. Both units have a total
    length of \SI{170}{\milli\meter}.}
    \label{fig:coils}
\end{figure}

\section{Energy harvester design}
Since the goal of this research is to highlight the viability and potential of
MFEH in railway, limited effort will be spent researching the optimal core
design---a substantial body of research that can be readily employed is
available on the subject of power grid solutions in the literature. A central
finding is that a ferrite core with a narrower diameter at its centre than at
its ends is beneficial in terms of efficiency \cite{yuan15,wright19}. In
general, the effective permeability $\mu_\mathrm{e}$ of a given design depends
on its geometry \cite{jiles16}, and according to the model
\Cref{eq:power_output}, output power scales with $\mu_\mathrm{e}$ squared.  In
fact, \cite{wright19} reports that employing a funnel-shaped ferrite core to
help guide the magnetic flux may improve the power density by an order of
magnitude compared to a similarly-sized rectangular coil. The designs in this
paper were developed with these results in mind.

In this project, two harvester coil designs were developed and tested both in
the laboratory and in the field. Pictured in \Cref{fig:coils}, their design
employs a dumbbell shape inspired by the bow-tie shape introduced in
\cite{yuan15} and the funnel core in \cite{wright19}. While \cite{yuan15} argues
that their bow-tie shape has a slightly higher effective permeability than a
dumbbell shape, the latter shape was chosen as it is a simpler shape to
construct and therefore results in a more economically viable solution.

The relevant parameters of the two coils are shown in \Cref{tab:cores}. Both
core designs in this paper feature steel disks at either end, connected by a
number of cylindrical ferrite rods. Arranged in a hexagonal pattern, Coil A is
comprised of seven ferrite rods while Coil B only requires three. The steel
disks have a diameter of \SI{50}{\milli\meter} and a thickness of
\SI{10}{\milli\meter}, and each ferrite rod a diameter of \SI{8}{\milli\meter}
and length of \SI{150}{\milli\meter}.

\begin{table}
    \setlength{\tabcolsep}{0.69em}
    \renewcommand{\arraystretch}{1.3}
    \caption{Main coil parameters of relevance.}
    \label{tab:cores}
    \centering
    \begin{tabular}{lS[table-format=1]cS[table-format=3]S[table-format=5]S[table-format=2.1]S[table-format=4]S[table-format=2.1]}
        \toprule
        {} & {Rods} & {Cost} & {$A$ [\si{\milli\meter\squared}]} & {$N$} & {$R$
        [\si{\kilo\ohm}]} & {$L$ [\si{\henry}]} & {$\mu_\mathrm{e}$}\\
        \midrule
        Coil A & 7 & \$80 & 590 & 80000 & 17.2 & 1000 & 23.5 \\
        Coil B & 3 & \$30 & 334 & 62000 &  9.2 &  500 & 31.3 \\ 
        \bottomrule
    \end{tabular}
\end{table}

\subsection{Magnetic characteristics}
The ferrite rod material is Ferroxcube's \textit{4B1}, which has a relative
permeability of \num{250} and a resistivity of \SI{1e5}{\ohm\meter}
\cite{ferroxcube}. This material was chosen for three reasons. Firstly, its
resistivity is very high, meaning that eddy current losses will be
substantially reduced. Secondly, its relative permeability is close to the
point at which the effective permeability becomes saturated. \cite{yuan15}
estimates the point of saturation to be around $\mu_\mathrm{r} = 400$ for a
similar core, and that any increase in the relative permeability beyond this
will lead to rapidly diminishing returns. Thirdly, the material is affordable
and available in a diverse range of shapes and sizes.

\begin{figure}
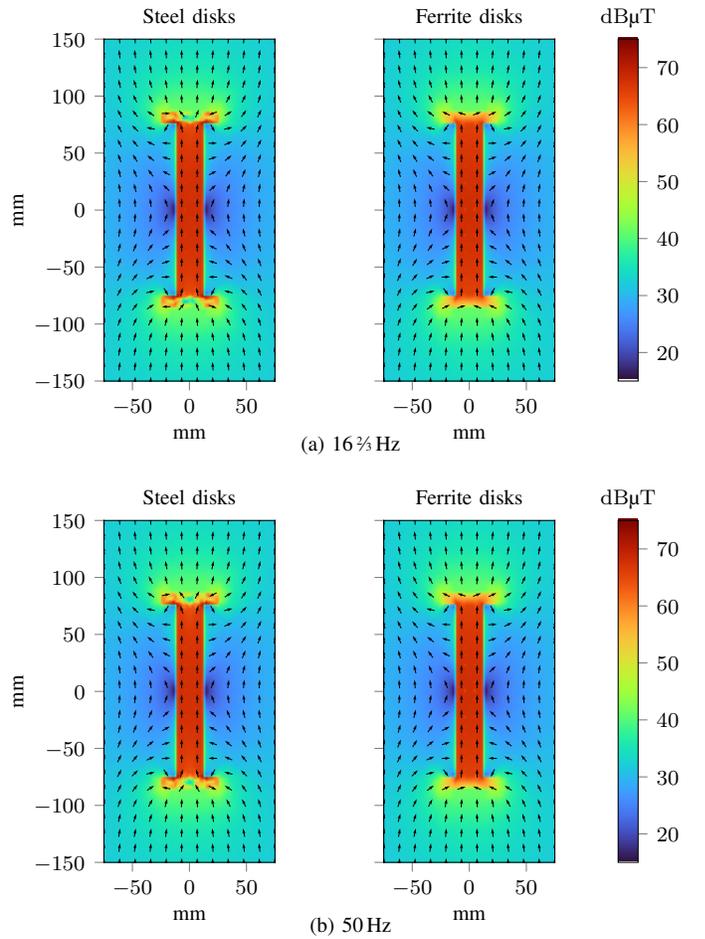

    \captionsetup[subfloat]{farskip=0pt}
    \captionsetup[subfloat]{captionskip=-2pt}
    \hfill
    \subfloat[\label{fig:core_field_16}\norrailfreq{}]{%
        \input{core_7r_field_16Hz.pgf}
    }
    \vspace{4mm}

    \hfill
    \subfloat[\label{fig:core_field_50}\SI{50}{\hertz}]{%
        \input{core_7r_field_50Hz.pgf}
    }
    \caption{The figures show the cross section of a 3D simulation of Core A.
    The negative effect of using steel disks as flux guides instead of ferrite
    is negligible in the weak, low-frequency magnetic fields encountered in
    railway, as illustrated by applying a uniform magnetic field of
    \SI{25}{\micro\tesla} RMS at \protect\subref{fig:core_field_16}
    \norrailfreq{} and \protect\subref{fig:core_field_50} \SI{50}{\hertz}. In
    \protect\subref{fig:core_field_16}, the magnetic flux $\varPhi_\mathrm{c}$
    through the core cross section at $y = 0$ is reduced by \SI{0.65}{\percent}
    if steel is employed in place of ferrite. In
    \protect\subref{fig:core_field_50}, the reduction is \SI{1.38}{\percent}.
    The results for Core B are of a similar nature.}
    \label{fig:core_7r_field}
\end{figure}

Steel disks were used at either end to further guide the magnetic flux towards
the ferrite cores. Theoretically, ferrite would be a more efficient material as
its higher resistivity would reduce eddy current loss. However, steel was
chosen for the flux guides as a compromise; the required shape was challenging
and expensive to acquire, and would therefore presumably be undesirable for an
end-user. Nonetheless, due to the narrow thickness of the disks and weak,
low-frequency magnetic fields involved, it is demonstrated in
\Cref{fig:core_7r_field} that eddy currents in the steel disks do not
substantially impact the magnetic characteristics of the cores.

Hysteresis loss is not likely to play a huge role in efficiency, as the
frequency and magnitude of the applied magnetic field are both very low. The
ferrite material has a magnetic loss tangent of less than \num{90e-6} when
applied a magnetic field of \SI{0.25}{\milli\tesla} at a frequency of
\SI{1}{\mega\hertz} \cite{ferroxcube}, and presumably even less when the
frequency is many orders of magnitude lower. It is therefore expected that the
material response is for all practical purposes linear and in phase with the
applied field.

The effective permeability listed in \Cref{tab:cores} was determined by
simulation as outlined in \cite{tashiro15}, using the expression
\begin{equation}\label{eq:eff_permeability}
    \mu_\mathrm{e} = \frac{V_\mathrm{core}}{V_\mathrm{air}} \ ,
\end{equation}
where $V_\mathrm{core}$ and $V_\mathrm{air}$ are the induced open-circuit RMS
voltages for the given coil and a corresponding air-core coil, respectively.

\subsection{Electrical characteristics}
Using enamelled copper wire with a diameter of \SI{0.1}{\milli\meter},
\num{80000} windings were applied to Coil A and \num{62000} windings to Coil B.
The core geometry and number of windings resulted in Coil A rendering an
inductance of \SI{1000}{\henry} and a resistance of \SI{17.2}{\kilo\ohm}, and
Coil B approximately halving both figures with an inductance of
\SI{500}{\henry} and a resistance of \SI{9.2}{\kilo\ohm}.

%% file: 5_lab_results.tex
\section{Laboratory results}\label{sec:lab_results}
An experiment was conducted in a smart grid laboratory in order to validate the
design and performance of the energy harvester coils, as well as to verify the
correctness of the theoretical model. The controlled environment allowed the
energy harvesting system to be tested at a range of distances, with currents of
configurable magnitudes and frequencies.

\subsection{Setup}
\begin{figure}[!t]
    \centering
    \input{lab_circ.tex}
    \caption{Equivalent circuit diagram of the laboratory setup. The rail
    current circuit (a) induces an open-circuit voltage $V_\mathrm{oc}$ in the
    energy harvester (b), from which the maximum power output can be derived.}
    \label{fig:lab_circ}
\end{figure}
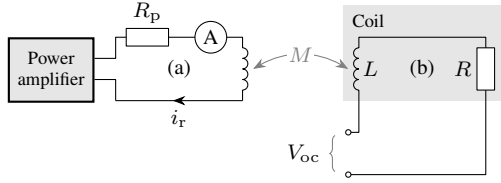

\begin{figure}
    \centering
    \includegraphics[width=\columnwidth]{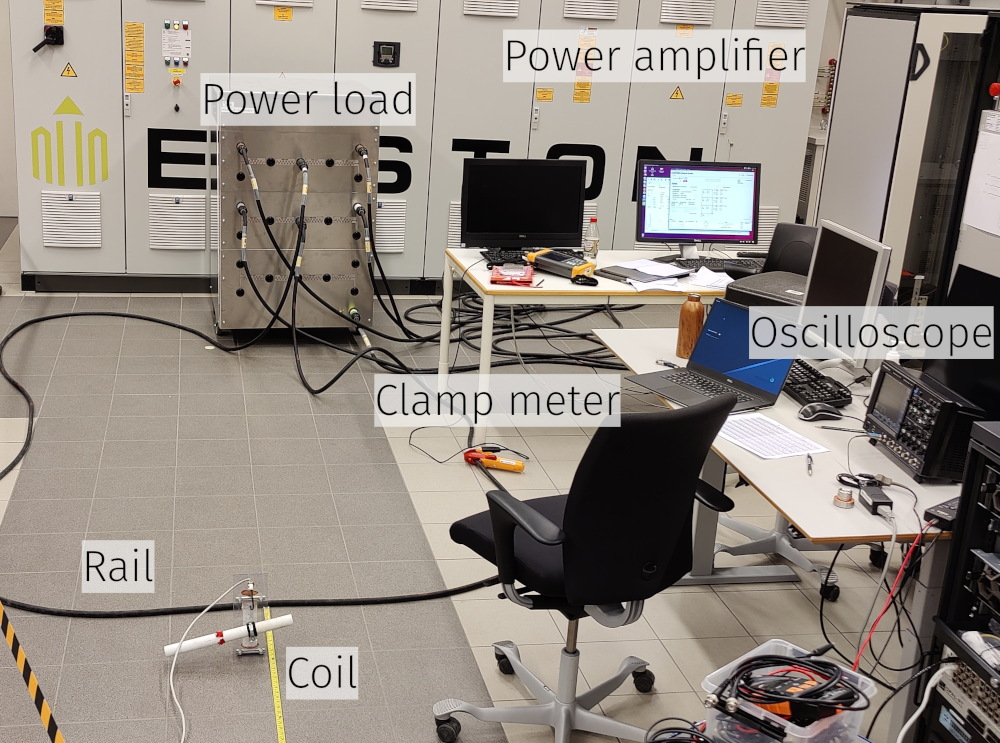}
    \caption{The physical setup used in the lab. The rail is emulated using a
    low-resistance cable.}
    \label{fig:lab_setup}
\end{figure}

As highlighted in the equivalent circuit in \Cref{fig:lab_circ}, the
experimental setup consisted of two galvanically isolated circuits: (a) a
high-current circuit representing the rail, which is magnetically coupled to
(b) the low-power energy harvesting circuit. While a real-world scenario will
have two parallel current-carrying rails separated by some distance, matters
were simplified in the experimental setup by only emulating a single rail. The
laboratory setup is depicted in \Cref{fig:lab_setup}.

\subsubsection{Rail current emulation}
The rail was realised as a low-resistance cable which carried a controllable AC
current, and thereby gave rise to the magnetic field from which energy could be
harvested. The configurable current of up to \SI{200}{\ampere} was generated by
a power amplifier (EGSTON CSU200), through a power resistor $R_\mathrm{p}$ of
\SI{800}{\milli\ohm}. A clamp meter (Fluke 325) was employed as a feedback
mechanism in order to ensure that the desired current magnitude was attained.

\subsubsection{Energy harvesting circuit}
The energy harvester was placed a precise distance from the high-current cable.
The induced open-circuit RMS voltage in the energy harvester was measured using
an oscilloscope (Teledyne Lecroy WaveJet 325) with an active differential probe
in order to eliminate any load on the circuit. From this measurement, $P_\ell$
was calculated as the output for a perfectly impedance-matched load with a
compensating capacitance.

\subsection{Model adaption}
The expression for the applied magnetic field $H_0$ as defined in
\Cref{sec:mag_field} will need to be adapted to the laboratory scenario, since
the assumption that the conductor is much longer than the distance to the coil
is not valid. In addition, the current runs as a loop, meaning that the reverse
polarity of the magnetic field generated from the far side needs to be
accounted for. The loop is fairly symmetric, so any contribution from the sides
can be assumed to cancel out.

To this end, the integral form of the Biot-Savart law \Cref{eq:biot_savart_int}
is used to derive the magnetic field at a distance $r$ from a conductor of
length $a$.
\begin{equation}\label{eq:biot_savart_int}
    \mathbf{H} = \frac{i_\mathrm{r}}{4\pi}\int\limits_{\mathcal{C}}
    \frac{\text{d}\boldsymbol{\ell} \times \hat{\mathbf{r}}}{|\mathbf{r}|^2}
\end{equation}
Using the substitutions
\begin{equation}\label{eq:subst}
    |\mathbf{r}|^2 = x^2 + r^2 \qquad \text{and} \qquad
    \text{d}\boldsymbol{\ell} \times \hat{\mathbf{r}} = \frac{r
    \hat{\boldsymbol{y}}}{\sqrt{x^2 + r^2}} \, \text{d}x \ ,
\end{equation}
an expression of the field can be derived as shown in \Cref{eq:biot_savart4}.
The resulting definition is similar to that in \Cref{eq:biot_savart2}, with an
additional correction factor depending on $a$ and $r$.
\begin{equation}\label{eq:biot_savart4}
    \mathbf{H} = \frac{i_\mathrm{r}}{4\pi} \int\limits_{-\sfrac{a}{2}}^{\sfrac{a}{2}}
    \frac{r\hat{\boldsymbol{y}}}{\left(x^2 + r^2\right)^{\sfrac{3}{2}}} \, \text{d}x =
    \frac{i_\mathrm{r}}{2\pi r} \frac{a}{\sqrt{4r^2 + a^2}} \hat{\boldsymbol{y}}
\end{equation}

To account for both the near and far side of the loop, two similar magnetic
fields are superimposed on each other. The resulting applied field $H_0$ is
shown in \Cref{eq:lab_field}, where $b$ is the distance between the far and near
side of the loop, measured to \SI{3}{\meter} in the laboratory.
\begin{equation}\label{eq:lab_field}
    H_0 = \frac{i_\mathrm{r}}{2\pi r} \frac{a}{\sqrt{4r^2 + a^2}} -
    \frac{i_\mathrm{r}}{2\pi (r + b)} \frac{a}{\sqrt{4(r + b)^2 + a^2}}
\end{equation}

\subsection{Results and discussion}

\begin{figure*}
    \captionsetup[subfloat]{farskip=0pt}
    \captionsetup[subfloat]{captionskip=-2pt}
    \subfloat[%
        \label{fig:lab_results_16}Output power at \norrailfreq{}%
    ]{\input{lab_results_16.tex}}%
    \hspace{1.3em}%
    \subfloat[%
        \label{fig:lab_results_50}Output power at \SI{50}{\hertz}%
    ]{\input{lab_results_50.tex}}%
    \caption{Expected and measured power output $P_\ell$ from Coils A and B as
    a function of rail current $I$, at \protect\subref{fig:lab_results_16}
    \norrailfreq{} and \protect\subref{fig:lab_results_50} \SI{50}{\hertz}.}
    \label{fig:lab_results}
\end{figure*}
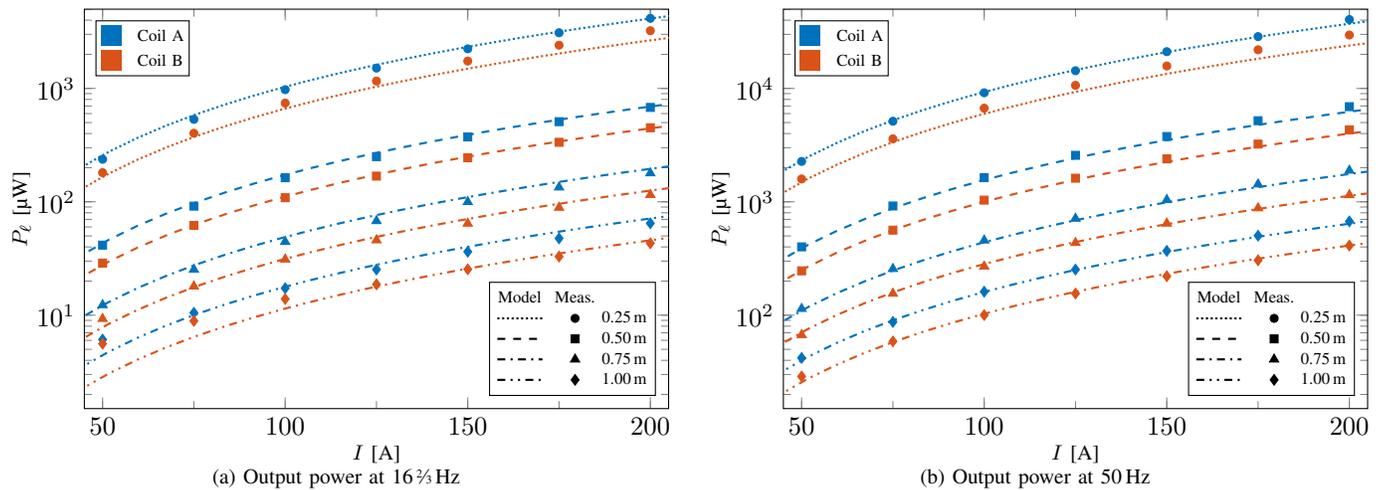

\begin{figure}
    \centering
    \includegraphics[width=\columnwidth]{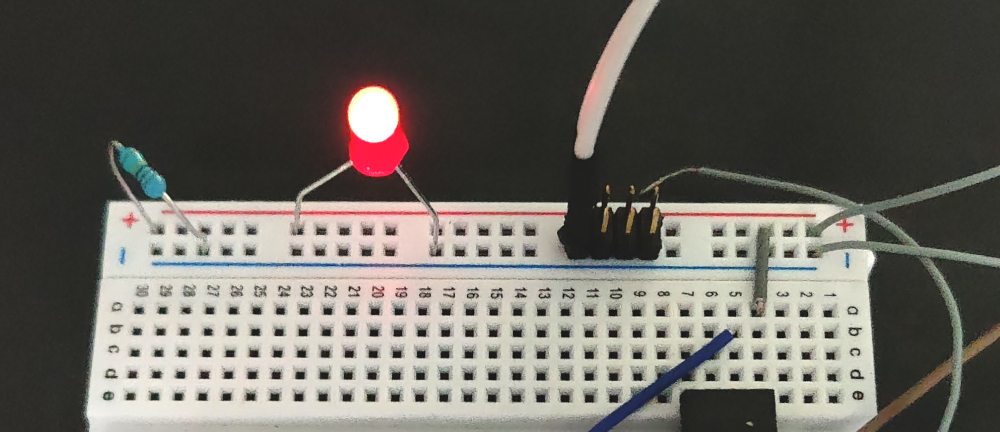}
    \caption{As an application example, the energy harvester can continuously
    power an LED. Here shown as a \SI{20}{\milli\watt} LED powered by Coil A,
    positioned \SI{0.25}{\meter} from a \SI{50}{\hertz} rail current of
    \SI{150}{\ampere}.}
    \label{fig:lab_led}
\end{figure}

The results from the laboratory trials are summarised in \Cref{fig:lab_results}
and compared to expected values obtained from the model,
\Cref{eq:power_output}. The two coils were tested at both \norrailfreq{}
(\Cref{fig:lab_results_16}) and \SI{50}{\hertz} (\Cref{fig:lab_results_50}),
with measurements of the harvester's open-circuit voltage recorded for a series
of increasing current levels ranging from \SI{50}{\ampere} to
\SI{200}{\ampere}. In order to paint a detailed picture of how this parameter
affects performance, distances of \num{0.25}, \num{0.50}, \num{0.75}, and
\SI{1.00}{\meter} were chosen as a reasonable range representative of what may
be permitted in a real-world scenario. For instance in Norwegian railway,
height restrictions placed on trackside equipment dictate that the harvester
cannot be placed closer than \SI{0.5}{\meter} from the rail, due to risk of
damage from equipment such as snowploughs. In other jurisdictions, the core may
be placed closer to the rails.

\subsubsection{Power and energy output}
The power draw of an electric locomotive is highly dependent on the mass it
pulls and the slope of the track. The typical current requirement of a
passenger train along a flat section of railway is reported to be over
\SI{100}{\ampere}, and several hundred amperes if the track is sloped
\cite{olsen20}. The best-case power output from the two energy harvester coils
was achieved at a distance of \SI{0.25}{\meter} from a current of
\SI{200}{\ampere}. At \norrailfreq{} the output was \SI{4.15}{\milli\watt} and
\SI{3.23}{\milli\watt} for Coils A and B, respectively, while increasing the
frequency to \SI{50}{\hertz} revealed a power output of \SI{40.5}{\milli\watt}
and \SI{29.6}{\milli\watt} for the coils.

The energy output will depend on the number of trains passing by, and the
duration for which the locomotives draw current of this magnitude. However, the
given power output convincingly demonstrates the viability of MFEH in railway.
In fact, the instantaneous power output is more than sufficient to power an LED
directly, as demonstrated in \Cref{fig:lab_led}. According to \cite{boyle16},
the typical power requirements of a node in a wireless sensor network is in the
range of tens to hundreds of microwatts in active operation, and tens of
nanowatts in sleep-mode. With this in mind, the laboratory experiments provide
convincing evidence that a WSN node could be solely powered by MFEH, especially
if employed in conjunction with energy accumulation and adaptive duty-cycling
techniques. Due to the scheduled nature of railway traffic, prediction models
utilising knowledge of train timetables could be especially beneficial. Further
examination of the energy output was conducted as part of the in situ testing,
discussed in \Cref{sec:rw_results}.

\subsubsection{Model evaluation}
In the following, the results will be discussed in the context of the
theoretical model \Cref{eq:power_output}.

\begin{equation*}
    P_\ell = \frac{\left(N A \mu_\mathrm{e} \mu_0 \omega H_0 \right)^2}{4R}
    \tag{\ref{eq:power_output} revisited}
\end{equation*}

As shown in the \Cref{fig:lab_results}, the model is quite accurate in its
prediction of the laboratory measurements. Nonetheless, there is a divergence
between predicted and measured values in the lower end of the range of power
outputs. This discrepancy is likely due to the substantial amount of
electromagnetic interference (EMI) emitted from the various equipment in the
laboratory, leading to measurements in the microwatt range being inflated. In
fact, the EMI alone induced an open-circuit RMS voltage of around
\SI{350}{\milli\volt} in both coils, offsetting the power figures by
approximately \SI{1.78}{\micro\watt} for Coil A and \SI{3.33}{\micro\watt} for
Coil B.

With the correction term introduced in \Cref{eq:lab_field}, the model is quite
precise in terms of how the power output $P_\ell$ scales with the distance $r$.
There is a minor discrepancy for the measurements at distance
\SI{0.25}{\meter}, especially for Coil B, which is most likely attributable to
the increasing curvature of the magnetic field. In \cite{espe20}, this
assumption was shown to be largely valid from about \SI{0.4}{\meter} and
outwards when employing two current-carrying rails. A single-rail scenario,
such as the one employed in the laboratory, will presumably highlight the
discrepancy to a greater degree.

In terms of other parameters, the measurements show that a doubling of the rail
current results in a fourfold increase in the output power. This resonates well
with the model, which would dictate a quadratic relationship between rail
current and power output. Furthermore, it is anticipated from the model that a
change in the frequency of the rail current should have a quadratic impact on
the output power. And, indeed, increasing the frequency threefold from
\norrailfreq{} to \SI{50}{\hertz} does manifest as a power output increase by a
factor of close to \num{9}. The relationship is not exact, with minor
deviations presumably caused by nonlinear phenomena not accounted for in the
model. One such effect is eddy current losses, an empirical model for which was
discussed in the previous \Cref{eq:eddy_loss}. This model describes how the
losses scale with frequency squared, which could explain some of the
deviations. Hysteresis losses and effects of mutual induction are also not
taken into account in the model and could have played a larger part than
expected.

%% file: lab_circ.tex
\tikzstyle{terminal} = [
    circle, draw, fill=white, minimum size=2pt, inner sep=0
]

\tikzstyle{resistor_v} = [
    rectangle, draw, minimum width=6pt, minimum height=16pt, fill=white
]

\tikzstyle{resistor_h} = [
    rectangle, draw, minimum width=16pt, minimum height=6pt, fill=white
]

\begin{tikzpicture}[font=\footnotesize]
    \node[rectangle, draw, thick, minimum width=32pt, minimum height=24pt,
    align=center, font=\scriptsize, fill=gray9]
    (current_source) at (0, 0) {Power\\amplifier};

    \draw[line join=bevel,
        decoration={markings, mark=at position 0.5 with {\centerarrow}}
    ] (current_source.north east) ++(0, -8pt) -| ++(8pt, 8pt) -- ++(12pt, 0)
    node[resistor_h] (r_p) {} ++(8pt, 0) -- ++(4pt, 0) -- ++(12pt, 0)
    node[circle, draw, minimum size=12pt, inner sep=0, fill=white] (clamp_meter)
    {A} ++(6pt, 0) -| ++(6pt, -4pt) arc(90:-90:2pt) arc(90:-90:2pt)
    coordinate (return_cur) arc(90:-90:2pt) arc(90:-90:2pt) -- ++(0, -4pt)
    edge[postaction=decorate] coordinate[pos=0.5] (i_r)
    ++(-48pt, 0) ++(-48pt, 0) |- ++(-8pt, 8pt);

    \node[below] at (i_r) {$i_\mathrm{r}$}; 
    \node[above, yshift=2pt] at (r_p) {$R_\mathrm{p}$}; 
    \node[yshift=12pt] at (i_r) {(a)};

    \coordinate (coil) at ($(return_cur) + (44pt, 8pt)$);

    \path[fill=gray9] (coil) ++(-6pt, 16pt) coordinate (label) -| ++(61pt, -36pt) -| cycle;

    \draw[line join=bevel] (return_cur) ++(44pt, -8pt) coordinate (circ_begin)
        arc(270:90:2pt) arc(270:90:2pt) coordinate (l_s) arc(270:90:2pt) arc(270:90:2pt)
        |- ++(48pt, 4pt) -- ++(0, -12pt)
        node[resistor_v] (r_s) {} ++(0, -8pt) |- ++(-52pt, -32pt)
        node[terminal] (v_oc_neg) {} ++(0, 16pt)
        node[terminal] (v_oc_pos) {} -| ++(4pt, 16pt);

    \draw[<->, color=gray5] ($(return_cur)+(4pt, 0)$) to[out=30, in=150]
        node[pos=0.5, fill=white, inner sep=0.5pt] {$M$} ($(l_s)-(4pt, 0)$);

    \draw[decorate, color=gray5, decoration={brace, amplitude=2pt, raise=5pt}]
        (v_oc_neg) -- (v_oc_pos) coordinate[pos=0.5] (v_oc_label);

    \node[below right] at (label) {\scriptsize Coil};
    \node[left, xshift=-7pt] at (v_oc_label) {$V_\mathrm{oc}$}; 
    \node[right, xshift=-2pt] at (l_s) {$L$}; 
    \node[left, xshift=-2pt] at (r_s) {$R$}; 

    \node[yshift=8pt, xshift=24pt] at (circ_begin) {(b)};

\end{tikzpicture}

%% file: lab_results_16.tex
\begin{tikzpicture}[trim axis right]
    \begin{axis}[
        scale only axis,
        width=\columnwidth - 3.1em,
        height=0.6\columnwidth,
        xlabel={\footnotesize $I$ [\si{\ampere}]},
        ylabel={\footnotesize $P_\ell$ [\si{\micro\watt}]},
        ylabel shift=-5pt,
        ymode=log,
        xlabel shift=-3pt,
        xmin=45, xmax=205,
        ymin=1.5, ymax=5000,
        legend pos=south east,
        legend style={
            nodes={scale=0.63, transform shape},
            at=({1, 0}),
            anchor=south east,
            legend cell align=left,
            outer sep=5pt,
            inner sep=2pt
        },
        legend columns=2,
        ytick={1, 10, 100, 1000, 10000},
        xtick={50, 100, ..., 200},
        minor xtick={50, 75, ..., 200}
    ]

    \addlegendimage{legend image with text={Model}}
    \addlegendentry{}
    \addlegendimage{legend image with text={Meas.}}
    \addlegendentry{}

    \addlegendimage{model, densely dotted}
    \addlegendimage{measured, mark=*}
    \addlegendimage{model, dashed}
    \addlegendimage{measured, mark=square*}
    \addlegendimage{model, dashdotted}
    \addlegendimage{measured large, mark=triangle*}
    \addlegendimage{model, dashdotdotted}
    \addlegendimage{measured large, mark=diamond*}

    \addlegendentry{}
    \addlegendentry{\SI{0.25}{\meter}}
    \addlegendentry{}
    \addlegendentry{\SI{0.50}{\meter}}
    \addlegendentry{}
    \addlegendentry{\SI{0.75}{\meter}}
    \addlegendentry{}
    \addlegendentry{\SI{1.00}{\meter}}

    \node[
        draw, fill=white, scale=0.63,
        anchor=north west, outer sep=7pt, inner sep=3pt
    ] at (rel axis cs: 0, 1)
        {\shortstack[l]{
            \colorbox{pgfblue}{\rule{0pt}{6pt}\rule{6pt}{0pt}}\quad Coil A \\
            \colorbox{pgfred}{\rule{0pt}{6pt}\rule{6pt}{0pt}}\quad Coil B
        }};

    \input{plot_data_16.tex}

    \end{axis}
\end{tikzpicture}

%% file: lab_results_50.tex
\begin{tikzpicture}[trim axis right]
    \begin{axis}[
        scale only axis,
        width=\columnwidth - 3.1em,
        height=0.6\columnwidth,
        xlabel={\footnotesize $I$ [\si{\ampere}]},
        ylabel={\footnotesize $P_\ell$ [\si{\micro\watt}]},
        ylabel shift=-5pt,
        ymode=log,
        xlabel shift=-3pt,
        xmin=45, xmax=205,
        ymin=15, ymax=50000,
        legend pos=south east,
        legend style={
            nodes={scale=0.63, transform shape},
            at=({1, 0}),
            anchor=south east,
            legend cell align=left,
            outer sep=5pt,
            inner sep=2pt
        },
        legend columns=2,
        ytick={10, 100, 1000, 10000, 100000},
        xtick={50, 100, ..., 200},
        minor xtick={50, 75, ..., 200}
    ]

    \addlegendimage{legend image with text={Model}}
    \addlegendentry{}
    \addlegendimage{legend image with text={Meas.}}
    \addlegendentry{}

    \addlegendimage{model, densely dotted}
    \addlegendimage{measured, mark=*}
    \addlegendimage{model, dashed}
    \addlegendimage{measured, mark=square*}
    \addlegendimage{model, dashdotted}
    \addlegendimage{measured large, mark=triangle*}
    \addlegendimage{model, dashdotdotted}
    \addlegendimage{measured large, mark=diamond*}

    \addlegendentry{}
    \addlegendentry{\SI{0.25}{\meter}}
    \addlegendentry{}
    \addlegendentry{\SI{0.50}{\meter}}
    \addlegendentry{}
    \addlegendentry{\SI{0.75}{\meter}}
    \addlegendentry{}
    \addlegendentry{\SI{1.00}{\meter}}

    \node[
        draw, fill=white, scale=0.63,
        anchor=north west, outer sep=7pt, inner sep=3pt
    ] at (rel axis cs: 0, 1)
        {\shortstack[l]{
            \colorbox{pgfblue}{\rule{0pt}{6pt}\rule{6pt}{0pt}}\quad Coil A \\
            \colorbox{pgfred}{\rule{0pt}{6pt}\rule{6pt}{0pt}}\quad Coil B
        }};

    \input{plot_data_50.tex}

    \end{axis}
\end{tikzpicture}

%% file: 6_rw_results.tex
\begin{figure}
    \centering
    \captionsetup[subfloat]{farskip=0pt}
    \subfloat[\label{fig:rw_setup}] {%
        \includegraphics[height=0.9\columnwidth]{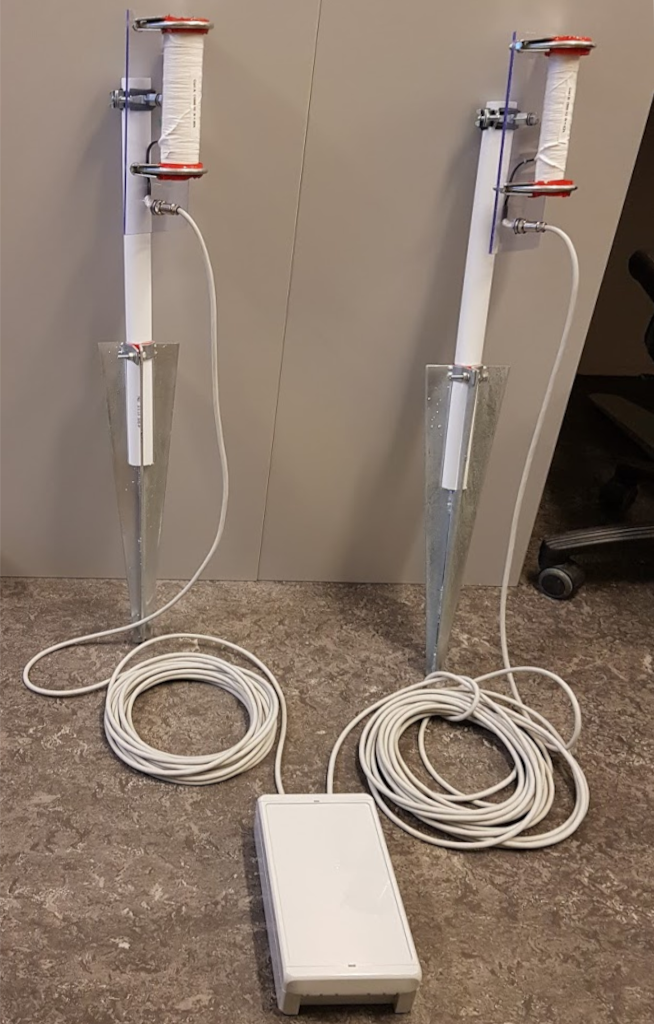}
    }
    \subfloat[\label{fig:map}] {%
        \includegraphics[height=0.9\columnwidth]{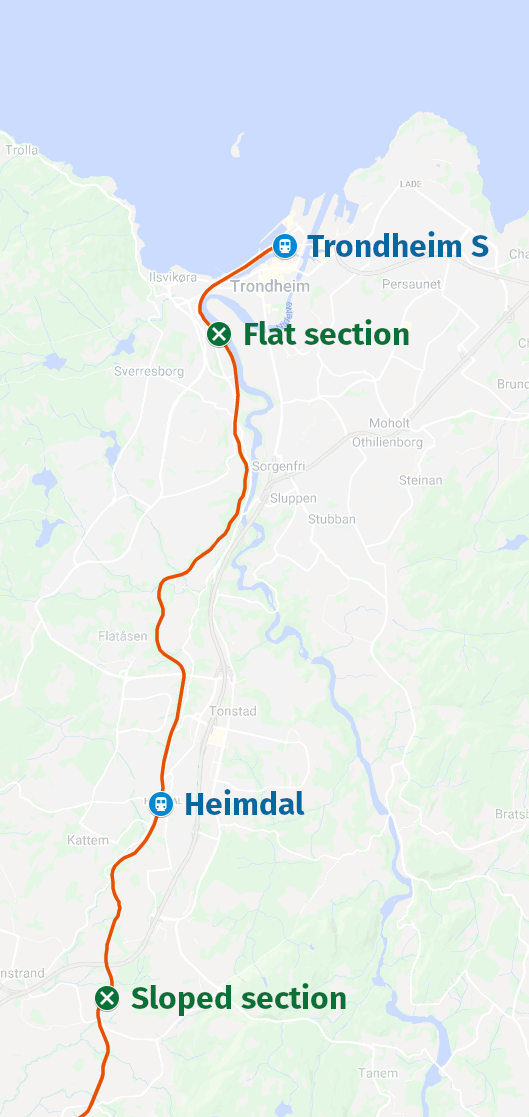}
    }
    \caption{The system to be placed in the field is shown in
    \protect\subref{fig:rw_setup}. In \protect\subref{fig:map}, the two test
    sites are marked. The flat and sloped sections are both along the Dovre
    line---\SI{550.52}{\kilo\metre} and \SI{537.64}{\kilo\metre},
    respectively.}
\end{figure}

\section{In situ results}\label{sec:rw_results}
In order to properly demonstrate the viability of MFEH in railway, the coils
were subsequently tested in a real-world scenario. The field setup, similar to
that of the lab tests, is depicted in \Cref{fig:rw_setup}. For practical
reasons, the coils were only tested along Norwegian railway, for which the
electrification frequency is \norrailfreq{}. Highlighted on the map in
\Cref{fig:map}, two different test sites were used: a flat section of railway
south of Trondheim Central station (Trondheim S), and an inclined section with
a slope of \SI{18}{\permille} (\SI{1.8}{\percent}) south of Heimdal station.

\begin{figure}
    \centering
    \includegraphics[width=\columnwidth]{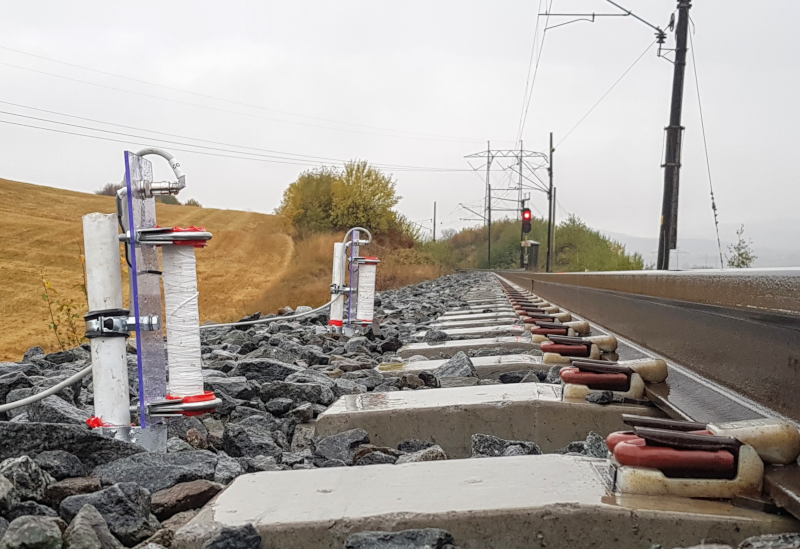}
    \caption{The coils were placed a distance of \SI{0.5}{\meter} from the
    nearest rail, here shown in the sloped location.}
    \label{fig:in-field}
\end{figure}

The installation procedure was completed quickly and effortlessly; only about
10 minutes passed from the point where access to the section of railway was
granted until the harvesting system was fully deployed and traffic could
resume. This speaks to the ease of installation and non-invasiveness of MFEH
along railway. Since access to the electrification system itself is not
necessary, a condition monitoring system employing MFEH can be quickly
installed with minimal impact on railway traffic. The installed system is
pictured in \Cref{fig:in-field}.

\subsection{Results and discussion}
The system was placed in each location for approximately two days, continuously
recording the power output from the energy harvesting coils. The envelope of
the output voltage indicates when a train draws current, and its magnitude is a
result of the amount of current drawn by the locomotive, and---due to ground
leakage currents---the locomotive's distance to the harvester coil. The signal
was matched with the train schedules to determine which train passed at which
point in time. Shown in \Cref{fig:recorded_data} are two representative
intervals from the recorded data, one from each test site.

\begin{figure*}
    \captionsetup[subfloat]{captionskip=-4pt}
    \captionsetup[subfloat]{farskip=0pt}
    \hfill
    \subfloat[\label{fig:rw_result_flat}Flat section] {%
        \input{rw_result_flat.pgf}%
    }%
    \hfill
    \subfloat[\label{fig:rw_result_sloped}Sloped section] {%
        \input{rw_result_sloped.pgf}%
    }%
    \caption{The closed-loop voltage $V_\ell$ across an impedance-matched load
    was recorded as trains passed by. The instantaneous dissipated power was
    derived as $P_\ell = \sfrac{V_\ell^2}{R_\ell}$. In
    \protect\subref{fig:rw_result_flat}, a passenger train followed a few
    minutes later by a freight train both depart from Trondheim S, drawing
    relatively short bursts of current. Chart
    \protect\subref{fig:rw_result_sloped} shows how a freight train sustains a
    larger current draw in order to maintain speed up the \SI{18}{\permille}
    slope.}
    \label{fig:recorded_data}
\end{figure*}

As illustrated in \Cref{fig:map}, the flat section is situated just
\SI{2.35}{\kilo\meter} south of Trondheim S, which means that trains running
past in either direction will not draw a substantial amount of current.
Northbound trains are arriving at the station and therefore generally do not
draw current, while southbound trains are leaving the station and limited to a
speed of \SI{30}{\kph}. According to the daily schedule, the trains recorded in
\Cref{fig:rw_result_flat} are both southbound trains leaving the station---one
passenger and one freight.

\Cref{fig:rw_result_sloped} shows the output from the energy harvester as a
northbound freight train passes by up the \SI{18}{\permille} slope. The rail
current is in the same order of magnitude as that seen for the flat section,
albeit maintained at higher levels for a considerably longer duration of time.
Compared to the short bursts of current in \Cref{fig:rw_result_flat}, a
substantial current is drawn for about 9 minutes. In this location, only
northbound trains, i.e. those running up the sloped section, will draw a
significant current. It is presumably locations like this in which MFEH will be
most effective.

\subsubsection{Power and energy output}
In \Cref{fig:rw_result_flat}, the voltage induced in Coil A by the passenger
train peaks at \SI{2}{\volt} amplitude, resulting in an instantaneous power
output of \SI{233}{\micro\watt}. For the freight train passing by a few minutes
later, the peak induced voltage is \SI{4.17}{\volt} and the power output
\SI{1.01}{\milli\watt}. For Coil B, the power figures are approximately
halved---\SI{169}{\micro\watt} and \SI{569}{\micro\watt} for the passenger and
freight trains, respectively.

Comparing \Cref{fig:rw_result_flat,fig:rw_result_sloped}, it is evident that
the peak voltages induced by the two freight trains are similar. However, the
power output from the sloped section is sustained for a substantially longer
time. Accumulating the power output as the freight train passes by the sloped
section gives a figure of \SI{109}{\milli\joule} and \SI{80.4}{\milli\joule}
for Coils A and B, respectively.

Typically, around nine northbound freight trains pass by this point every day,
which---through extrapolation---renders a daily energy harvest of around
\SI{981}{\milli\joule} and \SI{724}{\milli\joule} for Coils A and B,
respectively. When accounting for the four northbound electric passenger trains
that also pass by (\SI{40}{\milli\joule} and \SI{26}{\milli\joule} per train),
the estimates become \SI{1.14}{\joule} and \SI{828}{\milli\joule} per day for
Coils A and B. Both coils therefore seem to be viable solutions for powering
condition monitoring systems along railway. \cite{adumanu18} reports instances
of structural health monitoring systems requiring as little as
\SI{132}{\milli\joule} per day, a little less than one tenth of the amount
generated by Coil A on the sloped section.

The location of the energy harvester substantially impacts the amount of energy
that can be scavenged. It was shown that a steeper incline increases the power
output, although unless regenerative braking is used, only trains running up
the slope will provide a significant rail current. Apart from the slope, there
are several other factors at play. The current return path will generally be
directed towards the closest power substation, which means that only when the
harvester is located between the locomotive and the substation will there be a
substantial current in the rails. Placing the coil close to a power substation
will therefore presumably increase average power output.

It should be noted that for both in situ tests, the coils were placed
\SI{0.5}{\meter} from the rail due to restrictions placed on the height of
trackside equipment. If the coils were constructed in a more compact
form-factor, they could have been placed much closer to the rail, and---as seen
in the laboratory trials---provided a substantially larger power output.

%% file: 7_conclusion.tex
\section{Conclusions}
In this article, magnetic field energy harvesting (MFEH) in railway was
modelled, tested, and demonstrated in situ. Initially, models were derived for
the magnetic fields encountered in railway and the power output of an MFEH
device. Following this modelling effort, two prototype energy harvesting coils
were constructed, testing of which was performed in a controlled laboratory
environment to validate their designs and verify the model. In the laboratory,
the largest harvester rendered a power output of \SI{40.5}{\milli\watt} when
placed a distance of \SI{0.25}{\meter} from a \SI{50}{\hertz} rail current of
\SI{200}{\ampere}. Ultimately, both energy harvesters were placed in situ at
two points along Norwegian railway to record real-world performance. The
results make a convincing case for viability of MFEH in railway, with one coil
harvesting approximately \SI{0.1}{\joule} from a single freight train passing
by up a sloped section. Depending on the number of trains passing by, the
estimated daily output is in excess of \SI{1}{\joule}---a sufficient energy
budget for a low-power sensor node.

The power output was shown to be dependent on a number of factors, chief among
them the magnitude and duration of the rail current. The amount of energy that
can be harvested is therefore governed by the location of the system, the
railway topology, as well as the amount traffic. Other factors include the
coil's distance to the rail and the magnetic properties of the core. It is
expected that the power output may be substantially improved by increasing the
core's effective permeability, and by employing designs that allow placement
closer to the rails.

This article provides the first known in situ demonstration of MFEH in railway.
The concept lends itself to a quick, cost-effective, and non-invasive
installation procedure, and could therefore be an attractive approach for
railway administrations to extend the lifetime of battery-powered condition
monitoring systems. Furthermore, the magnetic fields encountered in railway
serve as a predictable source of energy, permitting power management
optimisation techniques to employ train schedules to further improve energy
consumption of trackside equipment.

%% file: paper.bbl
\begin{thebibliography}{10}
\providecommand{\url}[1]{#1}
\csname url@samestyle\endcsname
\providecommand{\newblock}{\relax}
\providecommand{\bibinfo}[2]{#2}
\providecommand{\BIBentrySTDinterwordspacing}{\spaceskip=0pt\relax}
\providecommand{\BIBentryALTinterwordstretchfactor}{4}
\providecommand{\BIBentryALTinterwordspacing}{\spaceskip=\fontdimen2\font plus
\BIBentryALTinterwordstretchfactor\fontdimen3\font minus
  \fontdimen4\font\relax}
\providecommand{\BIBforeignlanguage}[2]{{%
\expandafter\ifx\csname l@#1\endcsname\relax
\typeout{** WARNING: IEEEtran.bst: No hyphenation pattern has been}%
\typeout{** loaded for the language `#1'. Using the pattern for}%
\typeout{** the default language instead.}%
\else
\language=\csname l@#1\endcsname
\fi
#2}}
\providecommand{\BIBdecl}{\relax}
\BIBdecl

\bibitem{eu18}
{Publications Office of the European Union}, ``{EU transport in figures},''
  Luxembourg, 2018.

\bibitem{jbd18}
{Norwegian Railway Directorate}, ``{Jernbanestatistikk 2018},'' Oslo, Norway,
  2019, (in Norwegian).

\bibitem{akkermans16}
H.~Akkermans, L.~Besselink, L.~van Dongen, and R.~Schouten, ``{Smart Moves for
  Smart Maintenance},'' in \emph{World Class Maintenance}, 2016.

\bibitem{banenor17}
{Bane NOR (Norwegian National Rail Administration)}, ``{We create the railway
  of the future},'' Oslo, Norway, 2017.

\bibitem{takikawa16}
M.~Takikawa, ``{Innovation in railway maintenance utilizing information and
  communication technology (smart maintenance initiative).}'' \emph{{Japan
  Railway \& Transport Rev.}}, no.~67, pp. 22--35, 2016.

\bibitem{hodge15}
V.~J. Hodge, S.~O'Keefe, M.~Weeks, and A.~Moulds, ``{Wireless Sensor Networks
  for Condition Monitoring in the Railway Industry: A Survey},'' \emph{{IEEE}
  Trans. Intell. Transp. Syst.}, vol.~16, no.~3, pp. 1088--1106, Jun. 2015.

\bibitem{akyildiz02}
I.~F. Akyìldìz, W.~Su, Y.~Sankarasubramaniam, and E.~Çayırcı, ``{Wireless
  sensor networks: A survey},'' \emph{Comput. Netw.}, vol.~38, no.~4, pp.
  393--422, Mar. 2002.

\bibitem{boyle16}
D.~E. Boyle, M.~E. Kiziroglou, P.~D. Mitcheson, and E.~M. Yeatman, ``{Energy
  Provision and Storage for Pervasive Computing},'' \emph{{IEEE} Pervasive
  Comput.}, vol.~15, no.~4, pp. 28--35, Oct. 2016.

\bibitem{gjerstad20}
A.~Gjerstad, Bane NOR (Norwegian National Rail Administration), Aug. 2020,
  personal communication.

\bibitem{adumanu18}
K.~S. Adu-Manu, N.~Adam, C.~Tapparello, H.~Ayatollahi, and W.~Heinzelman,
  ``{Energy-Harvesting Wireless Sensor Networks (EH-WSNs): A Review},''
  \emph{{ACM} Trans. Sensor Netw.}, vol.~14, no.~2, pp. 1--50, Apr. 2018.

\bibitem{ulianov20}
C.~Ulianov, Z.~Hadaš, P.~Hyde, and J.~Smilek, ``{Novel Energy Harvesting
  Solutions for Powering Trackside Electronic Equipment},'' in
  \emph{{Sustainable Rail Transport}}, M.~Marinov and J.~Piip, Eds.\hskip 1em
  plus 0.5em minus 0.4em\relax Cham, Switzerland: Springer, 2020.

\bibitem{yang20}
F.~Yang, L.~Du, H.~Yu, and P.~Huang, ``{Magnetic and Electric Energy Harvesting
  Technologies in Power Grids: A Review},'' \emph{Sensors}, vol.~20, no.~5, p.
  1496, Apr. 2020.

\bibitem{kuang21}
Y.~Kuang, Z.~J. Chew, T.~Ruan, T.~Lane, B.~Allen, B.~Nayar, and M.~Zhu,
  ``{Magnetic field energy harvesting from the traction return current in rail
  tracks},'' \emph{{Applied Energy}}, vol. 292, pp. 1--14, Jun. 2021.

\bibitem{espe20}
A.~E. Espe and G.~Mathisen, ``{Towards Magnetic Field Energy Harvesting near
  Electrified Railway Tracks},'' in \emph{9th Mediterranean Conf. on Embedded
  Comput.}, Budva, Montenegro, 2020, pp. 1--4.

\bibitem{yuan15}
S.~Yuan, Y.~Huang, J.~Zhou, Q.~Xu, C.~Song, and P.~Thompson, ``{Magnetic Field
  Energy Harvesting under Overhead Power Lines},'' \emph{{IEEE} Trans. Power
  Electron.}, vol.~30, no.~11, pp. 6191--6202, Nov. 2015.

\bibitem{yuan17}
S.~Yuan, Y.~Huang, J.~Zhou, Q.~Xu, C.~Song, and G.~Yuan, ``{A High-Efficiency
  Helical Core for Magnetic Field Energy Harvesting},'' \emph{{IEEE} Trans.
  Power Electron.}, vol.~32, no.~7, pp. 5365--5376, Jul. 2017.

\bibitem{jiang16}
W.~Jiang, J.~Lu, S.~Hashimoto, and Z.~Lin, ``{A non-intrusive magnetic energy
  scavanger for renewable power generation state monitoring},'' in \emph{{IEEE}
  Int. Conf. on Renewable Energy Res. and Appl.}, Birmingham, UK, 2016, pp.
  562--566.

\bibitem{wu18}
Z.~Wu, D.~S. Nguyen, R.~M. White, P.~K. Wright, G.~O'Toole, and J.~R. Stetter,
  ``{Electromagnetic energy harvester for atmospheric sensors on overhead power
  distribution lines},'' \emph{J. of Phys.: Conf. Ser.}, vol. 1052, no.~1, pp.
  1--4, Jul. 2018.

\bibitem{pourghodrat14}
A.~Pourghodrat, C.~A. Nelson, S.~E. Hansen, V.~Kamarajugadda, and S.~R. Platt,
  ``{Power harvesting systems design for railroad safety},'' \emph{Proc. of the
  Institution of Mech. Engineers, Part F: J. of Rail and Rapid Transit}, vol.
  228, no.~5, pp. 504--521, Apr. 2014.

\bibitem{wischke11}
M.~Wischke, M.~Masur, M.~Kröner, and P.~Woias, ``{Vibration harvesting in
  traffic tunnels to power wireless sensor nodes},'' \emph{Smart Materials and
  Structures}, vol.~20, no.~8, p. 085014, Aug. 2011.

\bibitem{greenrail17}
\BIBentryALTinterwordspacing
{Community Research and Development Information Service}, ``{Greenrail,
  innovative and sustainable railway sleepers: the greener solution for railway
  sector },'' 2016, {Proj. No.: 738373---H2020-SMEINST-2-2016-2017}. [Online].
  Available: \url{https://cordis.europa.eu/project/id/738373}
\BIBentrySTDinterwordspacing

\bibitem{nandan17}
S.~Nandan, S.~Thakare, K.~Kulkarni, H.~Wagh, and G.~Magre, ``{T-BOX Wind Power
  Generation},'' in \emph{7th Intl. Conf. on Sci., Technol. \& Manage.},
  Nashik, India, 2017, pp. 148--156.

\bibitem{pan19}
H.~Pan, H.~Li, T.~Zhang, A.~A. Laghari, Z.~Zhang, Y.~Yuan, and B.~Qian, ``{A
  portable renewable wind energy harvesting system integrated S-rotor and
  H-rotor for self-powered applications in high-speed railway tunnels},''
  \emph{{Energy Conversion and Management}}, vol. 196, pp. 56--68, Sep. 2019.

\bibitem{kiessling18}
F.~Kiessling, R.~Puschmann, A.~Schmieder, and E.~Schneider, \emph{{Contact
  Lines for Electric Railways}}, 1st~ed.\hskip 1em plus 0.5em minus 0.4em\relax
  Munich, Germany: Publicis, 2001.

\bibitem{jbv97}
F.~Nilsen, ``{Sporstrømmer og potensialer i kontaktledningsanlegg},''
  Jernbaneverket, Tech. Rep., 1997, (in Norwegian).

\bibitem{ahmed19}
R.~Ahmed, B.~Buchli, S.~Draskovic, L.~Sigrist, P.~Kumar, and L.~T. Takikawa,
  ``{Optimal Power Management with Guaranteed Minimum Energy Utilization for
  Solar Energy Harvesting Systems},'' \emph{{ACM} Trans. on Embedded Comput.
  Syst.}, vol.~18, no.~4, p.~30, Jun. 2019.

\bibitem{lee20}
J.-H. Lee, J.-H. Kim, and Y.-G. Park, ``{Review of Minimum Curve Radius and
  Cant Range Setting for Mixed Section of Low and High speed Trains in
  Conventional Railway Line },'' \emph{{Journal of the Korea Academia --
  Industrial Cooperation Society}}, vol.~21, no.~10, pp. 345--353, Oct. 2020.

\bibitem{carmichael82}
R.~S. Carmichael, Ed., \emph{{Handbook of the Physical Properties of Rocks:
  Volume II}}.\hskip 1em plus 0.5em minus 0.4em\relax Boca Raton, FL, USA: CRC
  Press, 1982.

\bibitem{jiles16}
D.~Jiles, \emph{{Introduction to Magnetism and Magnetic Materials}},
  3rd~ed.\hskip 1em plus 0.5em minus 0.4em\relax Boca Raton, FL, USA: CRC
  Press, 2016.

\bibitem{roscoe13}
N.~M. Roscoe and M.~D. Judd, ``{Harvesting Energy from Magnetic Fields to Power
  Condition Monitoring Sensors},'' \emph{{IEEE} Sensors J.}, vol.~13, no.~6,
  pp. 2263--2270, Jun. 2013.

\bibitem{wright19}
S.~W. Wright, M.~E. Kiziroglou, S.~Spasic, N.~Radosevic, and E.~M. Yeatman,
  ``{Inductive Energy Harvesting From Current-Carrying Structures},''
  \emph{{IEEE} Sensors Lett.}, vol.~3, no.~6, pp. 1--4, Jun. 2019.

\bibitem{ferroxcube}
\BIBentryALTinterwordspacing
{Ferroxcube}, ``{4B1 Material Specification},'' 2008. [Online]. Available:
  \url{https://elnamagnetics.com/wp-content/uploads/library/Ferroxcube-Materials/4B1_Material_Specification.pdf}
\BIBentrySTDinterwordspacing

\bibitem{tashiro15}
K.~Tashiro, H.~Wakiwaka, and G.-Y. Hattori, ``{Estimation of Effective
  Permeability for Dumbbell-Shaped Magnetic Cores},'' \emph{{IEEE} Trans.
  Magn.}, vol.~51, no.~1, pp. 1--4, Jan. 2015.

\bibitem{olsen20}
B.~I. Olsen, Bane NOR (Norwegian National Rail Administration), Mar. 2020,
  personal communication.

\end{thebibliography}
